\documentclass[fleqn]{article}
\usepackage[a4paper, total={6.83in, 10in}]{geometry}
\usepackage{graphicx}
\usepackage{amsmath}
\usepackage{amsbsy}
\usepackage{epstopdf, epsfig}
\usepackage{amssymb}
\usepackage{longtable}
\usepackage{mathtools}
\usepackage{textcomp}
\usepackage{diagbox,color}
\usepackage{caption}
\title{\textbf{Mathematical Modeling, Laboratory Experiments, and Sensitivity Analysis of Bioplug Technology at Darcy Scale}}
\author{\textbf{David Landa-Marb\'an}, \textbf{Gunhild B{\o}dtker}, \textbf{Bartek Florczyk Vik}, and \textbf{Per Pettersson},\and Norwegian Research Centre; \textbf{Iuliu Sorin Pop}, University of Hasselt; \textbf{Kundan Kumar}, \and Karlstad University; and \textbf{Florin Adrian Radu}, University of Bergen}
\begin{document}
\maketitle
\section*{Summary}
In this paper we study a Darcy-scale mathematical model for biofilm formation in porous media. The pores in the core are divided into three phases: water, oil, and biofilm. The water and oil flow are modeled by an extended version of Darcy's law and the substrate is transported by diffusion and convection in the water phase. Initially there is biofilm on the pore walls. The biofilm consumes substrate for production of biomass and modifies the pore space which changes the rock permeability. The model includes detachment of biomass due to water flux and death of bacteria, and is implemented in MRST. We discuss the capability of the numerical simulator to capture results from laboratory experiments. We perform a novel sensitivity analysis based on sparse-grid interpolation and multi-wavelet expansion to identify the critical model parameters. Numerical experiments using diverse injection strategies are performed to study the impact of different porosity-permeability relations in a core saturated with water and oil. 

\section*{Introduction}
After primary and secondary production, up to 85\% of the oil remains in the reservoir~\cite{Patel:Article:2015}. Microbial improved and enhanced oil recovery (MIEOR) is one of the secondary and tertiary methods to increase the oil production using microorganisms~\cite{Wood:Article:2019}. Bioplug technology is a MIEOR strategy that consists in plugging the most permeable zones in the reservoir, which provokes water to flow through new paths and recovering the oil in these new zones.

Experiments in microsystems allow us to observe processes in more detail, which leads to improvement of the experimental methods in core-scale experiments. For example, in \cite{Liu:Article:2019} the effects of flow velocity and substrate (also referred to as nutrients) concentration on biofilm in a microchannel was studied, finding values of substrate concentration and flow velocity for a strong plugging effect. Core samples from reservoirs can be used to study changes in permeability due to biofilm formation, e.g., in~\cite{Suthar:Article:2009} two-phase flow experiments were performed to study the selective plugging strategy for MIEOR. In that study, the MIEOR effects increased the oil recovery \mbox{around 25\%.}    

Mathematical models of bioplug technology are important as they help to predict the applicability of this MIEOR strategy and to optimize the benefits. In \cite{Kim:Article:2006} a mathematical model for single-phase flow was proposed which includes changes of rock porosity and permeability as a result of biofilm growth. Li et al. \cite{Li:Article:2011} built a mathematical model for two-phase flow including the effects of bio-surfactants and biomass on improving the oil recovery. The authors also compare the numerical results  for different porosity-permeability relations. These porosity-permeability relations can also include the permeability of biofilm and be derived as a result of upscaling pore-scale models~\cite{vanNoorden:Article:2010,Hommel:Article:2018,Landa:Article:2019}. In this work, we present a two-phase core-scale model of bioplug technology and study the oil production for different porosity-permeability relations and injection strategies.  

Sensitivity studies of mathematical models are of great interest because they provide estimates of the influence of physical parameters on a quantity of interest, e.g., biofilm formation. In \cite{Brockmann:Article:2006} a regional steady-state sensitivity analysis was performed to identify parameters with the largest impact on a mathematical model for deammonification in biofilm systems. Sensitivity analysis by means of Sobol decomposition provides rigorous estimates of parameter dependencies, but are prohibitively expensive to compute if the number of parameters is large. This is remedied for smooth problems by first computing spectral (generalized polynomial chaos) expansions in the parameters, which then leads to efficient evaluation of the sensitivity indices via post-processing of spectral coefficients~\cite{Sudret_08}. The latter method was employed in~\cite{Landa:Article:2019a}, where a global sensitivity analysis was performed using Sobol indices to identify the critical parameters of a pore-scale model for permeable biofilm. In this paper, we consider nonsmooth models in the dependent parameters, hence spectral expansions with global smooth basis functions are not a robust choice. Instead, we propose a two-stage method where we first use sparse grids to estimate a piecewise linear interpolant of the function of interest, which yields a surrogate that can be further sampled at negligible cost. Secondly, we compute a multi-wavelet representation of the interpolant, from which the sensitivities can be directly evaluated. 

The mathematical model consists of coupled nonlinear partial differential equations. Two-point flux approximation (TPFA) and backward Euler (BE) are used for the space and time discretization respectively. To solve the resulting nonlinear algebraic system, Newton's method is used. The scheme is implemented in the MATLAB reservoir simulation tool (MRST), a free open-source software for reservoir modeling and simulation~\cite{Lie:Book:2019}.  

To summarize, the new contributions of this work are:
\begin{itemize}
\item The comparison of experimental results to numerical simulations of a core-scale porous medium including biofilm formation.
\item Performing a novel global sensitivity analysis of the model parameters to identify the critical parameters.
\item Performing numerical simulations for different porosity-permeability relations to study their impact on the predicted oil recovery. 
\item The study of diverse substrate injection strategies for the bioplug technology.
\end{itemize}
The paper is structured as follows. Firstly, we introduce and describe the implementation of the core-scale mathematical model for two-phase flow including the effects of biofilm formation. Secondly, we present a comparison of numerical simulations to laboratory experiments of this core-scale model. Thirdly, we introduce a novel method for global sensitivity analysis and apply it to the mathematical core-scale model. Diverse numerical experiments for different porosity-permeability relations and injection strategies are also explained. Finally, we present the conclusions.   

\section*{Core-Scale Model}\label{Core-scale model}
\begin{figure}[h!]
\includegraphics[width=\textwidth]{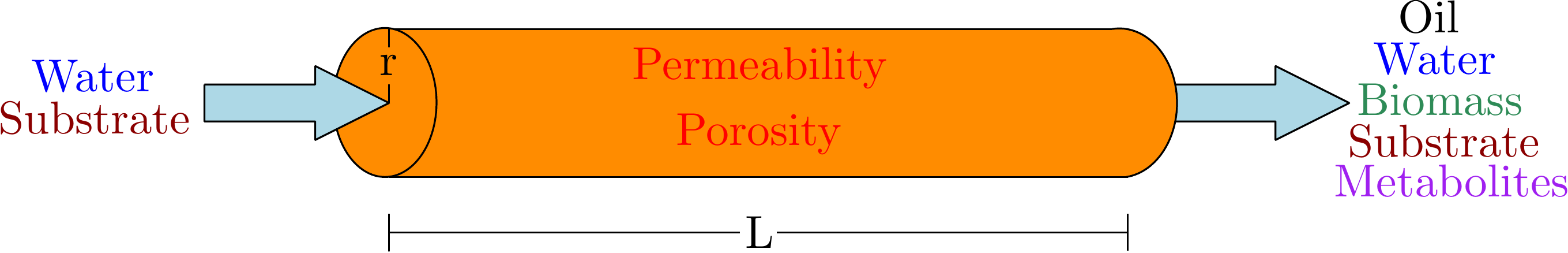}
\caption{---Core sample for laboratory experiments after inoculation of bacteria.}
\label{modelco} 
\end{figure}

We consider a core sample of radius $r$ and length $L$ initially filled with oil and water and a given biofilm distribution. \textbf{Fig. \ref{modelco}} shows the schematic representation of this system. As water and substrate are injected, some biomass is detached due to erosion (shear forces caused by the water flow). The biofilm consumes substrate to produce metabolites (i.e., gases) and to grow which modifies the rock porosity and hence the rock permeability. Consequently, the flow pattern is modified in the sense that pores where oil was replaced by water, and which were forming a preferential path for water flow, become less permeable. Therefore water enters other pores, mobilizing the oil present there and leading to improved oil production. The mathematical model presented in this section aims to describe the following processes in a core after bacterial inoculation: two-phase flow, substrate transport, permeability changes due to porosity modification, and biofilm growth, detachment, and death. 

\noindent We assume that the fluids are immiscible and incompressible. We consider that the biofilm has water and biomass as the only two components, and that the biofilm porosity and biofilm permeability are constants. In order to determine the amount of fluid outside the biofilm in the representative element volume (REV), we introduce the saturation of a fluid $S_\alpha$(for oil $\alpha=o$ and for water $\alpha=w$) given by the ratio of volume of fluid $\alpha$ outside the biofilm over the volume of voids outside the biofilm (in REV). For this two-phase flow model, we have that $S_o+S_w=1$. 

The mass conservation and extended Darcy's law for the oil phase are given by 
\begin{equation}\label{tpf}
\frac{\partial}{\partial t}(\rho_o\phi_f S_o)+\nabla\cdot(\rho_o\vec{v}_o)=Q_o,\quad\vec{v}_o=-\frac{k_{r,o}}{\mu_o}k(\nabla p_o-\rho_o\vec{g}),\quad\ldots\ldots\ldots\ldots\ldots\ldots\ldots\ldots\ldots\ldots\ldots\ldots\ldots\ldots\ldots
\end{equation}
and for the water phase 
\begin{equation}\label{tpfw}
\frac{\partial}{\partial t} [\rho_w(\phi_f S_w+\phi_b\theta_w)]+\nabla\cdot(\rho_w\vec{v}_w)=Q_w,\quad\vec{v}_w=-\frac{k_{r,w}}{\mu_w}k(\nabla p_w-\rho_w\vec{g}),\quad\ldots\ldots\ldots\ldots\ldots\ldots\ldots\ldots\ldots\ldots\ldots
\end{equation}
where $\phi_f$ is the porosity outside the biofilm, $\vec{v}_\alpha$ the flow velocity, $k_{r,\alpha}$ the relative permeability, $\rho_\alpha$ the fluid density, $k$ the absolute rock permeability, $\vec{g}$ the gravity, $\mu_\alpha$ the viscosity, $\phi_b$ the volume fraction of biofilm in the REV, $\theta_w$ the biofilm water content, and $Q_\alpha$ source/sink terms. The minimum value of $\phi_b$ is zero while the maximum value of $\phi_b$ is equal to the initial porosity of the rock $\phi_0$, corresponding to the case when the pores are filled with biofilm. In this work, we assume that there are no fluid sources/sinks, neglect the gravity effects, set to zero the residual oil saturation and irreducible water saturation, assume oil cannot penetrate into the biofilm, and neglect the capillary pressure $(p_w=p_o)$. The previous assumptions are commonly used to reduce the complexity and number of parameters of mathematical models for porous media. Consider the mass conservation for the water in Eq. \ref{tpfw}. The first term gives the changes on time for the total water mass in the REV, i.e., the water outside the biofilm $(\rho_w\phi_f S_w)$ plus the water inside the biofilm $(\rho_w\phi_b\theta_w)$. The Brook-Corey relations are commonly used to model the relative permeability relations. These relations are given as a function of the water saturation and experimental parameters such as exponents for the saturation, values for the endpoint relative permeabilities, and residual phase saturations. Simplified expressions of these relations are given by
\begin{equation}
k_{r,w}(S_w)=S_w^{\gamma},\quad k_{r,o}(S_w)=(1-S_w)^{\beta},\quad\ldots\ldots\ldots\ldots\ldots\ldots\ldots\ldots\ldots\ldots\ldots\ldots\ldots\ldots\ldots\ldots\ldots\ldots\ldots\ldots\ldots
\end{equation}
where $\gamma$ and $\beta$ are experimentally fitted factors. We assume that water can flow inside the biofilm even when there is only oil outside the biofilm ($S_w=0$). Then, the previous relation is not a good model because for $S_w=0$ it results in $k_{r,w}(0)=0$ which leads to zero water velocity. Recalling that the relative permeabilities are used to extend Darcy's law to multiphase flow, we look for relationships which account for the water saturation and volume fraction of biofilm. These relationships need to fulfill the following criteria:  when there is only oil outside the biofilm, then $k_{r,w}(S_w=0,\phi_b)=1$; when there is only biofilm in the REV, then $k_{r,w}(S_w,\phi_b=\phi_0)=1$. Then, we propose the following two relative permeability relations:
\begin{equation}
k_{r,w}(S_w,\phi_b)=S_w^{\gamma}\left (1-\frac{\phi_b}{\phi_0}\right)+\left (\frac{\phi_b}{\phi_0}\right),\quad k_{r,o}(S_w,\phi_b)=(1-S_w)^{\beta}\left (1-\frac{\phi_b}{\phi_0}\right).\quad\ldots\ldots\ldots\ldots\ldots\ldots\ldots\ldots\ldots
\end{equation}
To describe the movement of substrate, we consider the following convection-diffusion-reaction transport equations:
\begin{equation}
\frac{\partial}{\partial t} \left[C_n(\phi_f S_w+\phi_b\theta_w)\right]+\nabla\cdot \vec{j}_n=R_n,\quad \vec{j}_n=-D_n(\phi_f S_w+\phi_b\theta_w)\nabla C_n+C_n\vec{u}_w.\quad\ldots\ldots\ldots\ldots\ldots\ldots\ldots\ldots\ldots
\end{equation}
In the previous equations, $C_n$ is the substrate concentration in the water, $\vec{j}_n$ the substrate flux in water, and $D_n$ the substrate dispersion coefficient which includes mechanical dispersion plus diffusion. The reaction term $R_n$ is given by
\begin{gather}\label{rates}
\begin{split}
R_n=-\mu_n\frac{C_n}{K_n+C_n}\rho_b\phi_b,\quad\ldots\ldots\ldots\ldots\ldots\ldots\ldots\ldots\ldots\ldots\ldots\ldots\ldots\ldots\ldots\ldots\ldots\ldots\ldots\ldots\ldots\ldots\ldots\ldots\ldots
\end{split}
\end{gather} 
where $\mu_n$ is the maximum rate of substrate utilization, $\rho_b$ the biomass density, and $K_n$ is the Monod half-velocity coefficient. 

The following equation describes the biofilm evolution:
\begin{equation}
\frac{\partial}{\partial t}(\rho_b\phi_b)=\mu_n\frac{C_n}{K_n+C_n}\rho_b\phi_b-K_{d}\rho_b\phi_b-K_{str}||\nabla p|| \rho_b\phi_b.\quad\ldots\ldots\ldots\ldots\ldots\ldots\ldots\ldots\ldots\ldots\ldots\ldots\ldots\ldots\ldots\ldots
\end{equation}
Here, we assumed that all consumed substrate is used to produce biomass, a linear death of bacteria given by $K_{d}$, and erosion of bacteria given by the magnitude of the pressure gradient times a constant depending on the biofilm $K_{str}$. After biomass has been detached, in this model we assume that the detached biomass flows out of the core and does not affect the rock properties; therefore, we do not include a transport equation for the detached biomass. 

Diverse porosity-permeability relations have been proposed for the last decades. We refer to \cite{Hommel:Article:2018} for a recent review of these relations. Three porosity-permeability relations commonly used in modeling are the following:
\begin{equation}\label{Kp}
\frac{k_p}{k_0}=\bigg (\frac{\phi_f}{\phi_0}\bigg )^\eta,\;\frac{k_{vp}}{k_0}=\bigg (\frac{\phi_f-\phi_{crit}}{\phi_0-\phi_{crit}}\bigg )^\eta,\;\frac{k_{h}}{k_0}=a\bigg (\frac{\phi_f-\phi_{crit}}{\phi_0-\phi_{crit}} \bigg )^3+(1-a)\bigg (\frac{\phi_f-\phi_{crit}}{\phi_0-\phi_{crit}} \bigg )^2.\quad\ldots\ldots\ldots\ldots\ldots\ldots
\end{equation}
The first one is called the power law~\cite{Carman:Article:1937,Ives:Article:1965}, where $\eta$ is a fitting factor calibrated either from experimental data or taken from a process-specific literature. The second relation is a variation of the power law known as the Verma-Pruess relation~\cite{Verma:Article:1988}, where $\phi_{crit}$ is a critical porosity when the permeability becomes zero, which value is between 70 and 90\% of the initial porosity. The third relation is proposed by~\cite{Thullner:Article:2002}, where $a$ is a weighting factor between -1.7 and -1.9. These three relations do not include the biofilm permeability. Vandevivere \cite{Vandevivere:Article:1995} proposes the following relation of permeability and porosity for a plugging model
\begin{equation}\label{Kv}
\frac{k_v}{k_0}=\exp\left[-\frac{1}{2}\left(\frac{B_r}{B_c}\right )^2\right]\left(\frac{\phi_f}{\phi_0}\right)^2+\left\lbrace1-\exp\left[-\frac{1}{2}\left(\frac{B_r}{B_c}\right )^2\right]\right\rbrace\frac{k_b\phi_0}{k_0\phi_0-(k_0-k_b)\phi_f},\quad\ldots\ldots\ldots\ldots\ldots\ldots\ldots
\end{equation}
where $k_b$ is the biofilm permeability, $B_r$ is a relative porosity given by $B_r=1-\phi_f/\phi_0$, and $B_c$ is the critical point where biofilm begins to detach and form plugs. Thullner et al. \cite{Thullner:Article:2002} presents the following relation which includes the biofilm permeability:
\begin{equation}\label{Kth}
\frac{k_{th}}{k_0}=\bigg[\bigg(\frac{\phi_f-\phi_{crit}}{\phi_0-\phi_{crit}}\bigg)^\eta+\frac{k_b}{k_0}\bigg]\frac{k_0}{k_0+k_b}.\quad\ldots\ldots\ldots\ldots\ldots\ldots\ldots\ldots\ldots\ldots\ldots\ldots\ldots\ldots\ldots\ldots\ldots\ldots\ldots\ldots\ldots
\end{equation}

\noindent All these relationships are postulated at the Darcy-scale, based on experimental observations. A different approach is followed in \cite{vanNoorden:Article:2010}, \cite{Schulz:Article:2017}, and \cite{Landa:Article:2019}, based on homogenization. Starting with models at the pore scale, one applies expansion methods to derive the mathematical models valid at the Darcy scale. In~\cite{Landa:Article:2019} effective porosity-permeability relations are derived for two different pore geometries: thin channels $k_c$ and tubes $k_t$. In this work we compare the following four porosity-permeability relations: Vandevivere $k_v$ (\ref{Kv}), Thullner $k_{th}$ (\ref{Kth}), channel $k_c$, and tube $k_t$. We refer to Appendix A for the mathematical expressions of $k_c$ and $k_t$. 

The porosity in the porous medium changes in time as a function of the biofilm volume fraction $\phi_b(t)$. When there is no biofilm, the porosity $\phi_f(t)$ is equal to the initial porosity. In the case when the porous medium is filled with biofilm ($\phi_f(t)=0$), the porosity in the REV is equal to the biofilm porosity $\theta_w$. The porosity-permeability relations (\ref{Kv}-\ref{Kth}) are given as a function of the rock porosity and do not include the biofilm porosity. The porosity-permeability relations $k_c$ and $k_t$ do include the biofilm porosity. Using the definitions of $\phi_f(t)$ and $\phi_b(t)$, we have the following relation for the void space outside the biofilm
\begin{equation}
\phi_f(t)=\phi_0-\phi_b(t).\quad\ldots\ldots\ldots\ldots\ldots\ldots\ldots\ldots\ldots\ldots\ldots\ldots\ldots\ldots\ldots\ldots\ldots\ldots\ldots\ldots\ldots\ldots\ldots\ldots\ldots\ldots\ldots
\end{equation}
This mathematical model is implemented in a 3D domain with a uniform cell-centered grid. The TPFA is used for the spatial discretization, while BE for the time discretization. The Newton's method is used as linearization scheme. The implementation of this core-scale mathematical model is done in MRST, based on the modification of the polymer example in the module ad-eor (see \cite{Bao:Article:2017}). 

\section*{Model Test}\label{Model test}
\begin{figure}[h!]
\includegraphics[width=\textwidth]{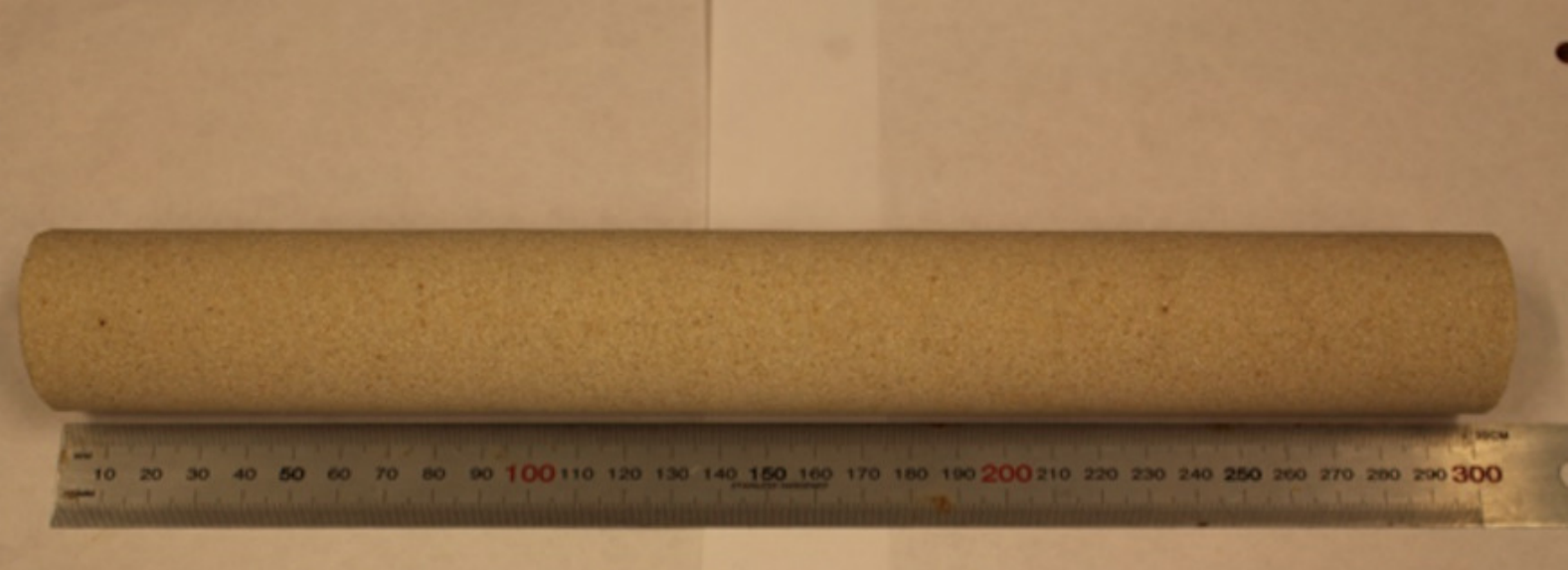}
\caption{---Core sample for laboratory experiments.}
\label{corepic} 
\end{figure}

\noindent Core-scale experiments under controlled conditions are performed in the laboratory for studying the effect on biofilm growth in porous media. These experiments aim to provide a better understanding of different features, i.e., the relation between biofilm composition and growth conditions, the plugging potential of different bacteria, and the adaptability of biofilm at diverse substrate flux. \textbf{Fig. \ref{corepic}} shows a photograph of a typical core sample to study these effects.\\

\begin{figure}[h!]
\includegraphics[width=\textwidth]{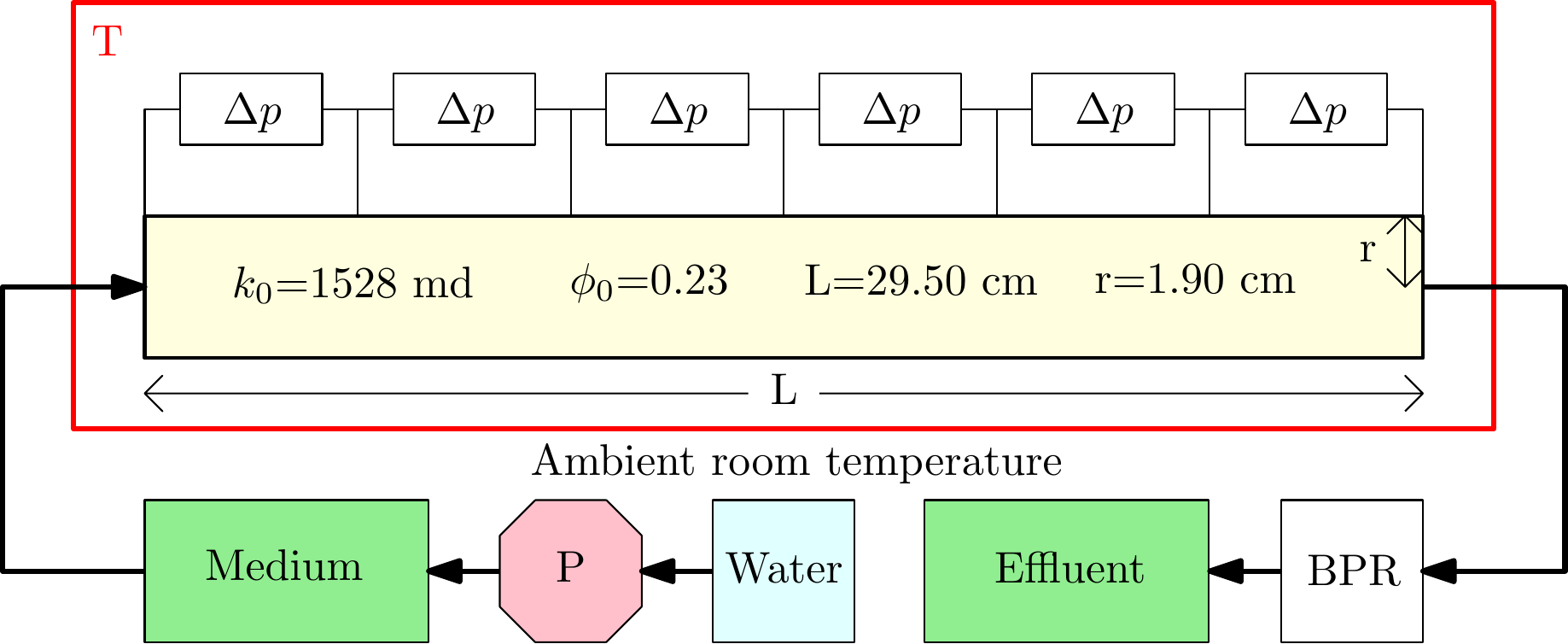}
\caption{---Scheme of an experimental set up for laboratory experiments.}
\label{coreschem} 
\end{figure}

\noindent In this section, we describe an experiment performed by NORCE for Equinor. The aim of this experiment is to test different brine qualities on the same system and the same biofilm. The core has a length of 29.50 cm and a radius of 1.90 cm. Initially, the core has an approximately homogeneous porosity of $\phi_0$=0.23 and permeability $k_0$=1528 md. The core sample is introduced inside a core holder, fully saturated with brine. Heating cables wrapped around the exterior of the core holder are used to conduct the experiments at a constant temperature (30\textdegree{}C). A backpressure regulator (BPR) is used to control the outlet pressure. Bacteria are injected into the core and the core is left standstill overnight to allow the bacteria to attach themselves to the pore walls. \textbf{Fig. \ref{coreschem}} shows a diagram of the experimental set-up.

\noindent The core is injected with two different brines with different substrate concentrations, where glucose is added as a source of carbon. We refer to these brines as B$_1$=$30.81$ kg/m$^3$ and B$_2$=$42.63$ kg/m$^3$. The two different brines are injected at a constant flow rate of 8.33$\times10^{-10}$ $m^3$/s (Darcy velocity of 6.4 cm/day). The injection strategy is the following: B$_1$ is injected during 53 days, after B$_2$ is injected during 82 days and finally B$_1$ is injected again during 61 days. The core permeability is estimated at different times, using the measurement of the pressure difference along the core divided by the pressure drop on the clean core. This ratio is known as resistance factor $R_f$ and it is given by $R_f(t)=\Delta p_w(t)/\Delta p_0$. Assuming that there is not change in the fluid rate, viscosity, and density, the resistance factor gives an estimate of the current rock permeability $k(t)=R_f(t)k_0$.

In the previous section we introduced the model equations for biofilm growth and two-phase flow in porous media. We simplify the mathematical model to compare to the experiment. This experiment can be modeled as a 1D single-phase flow system, where in addition we consider that the biofilm is impermeable and only diffusion for the substrate to reduce the number of parameters. \textbf{Table \ref{tab:2}} presents this simplified version of the mathematical model which includes only six variables: the water velocity $v_w$, water pressure $p_w$, permeability $k$, porosity $\phi$, substrate concentration $C_n$, and volume fraction of biomass $\phi_b$. 

\begin{figure}[h!]
\centering
\includegraphics[width=\textwidth]{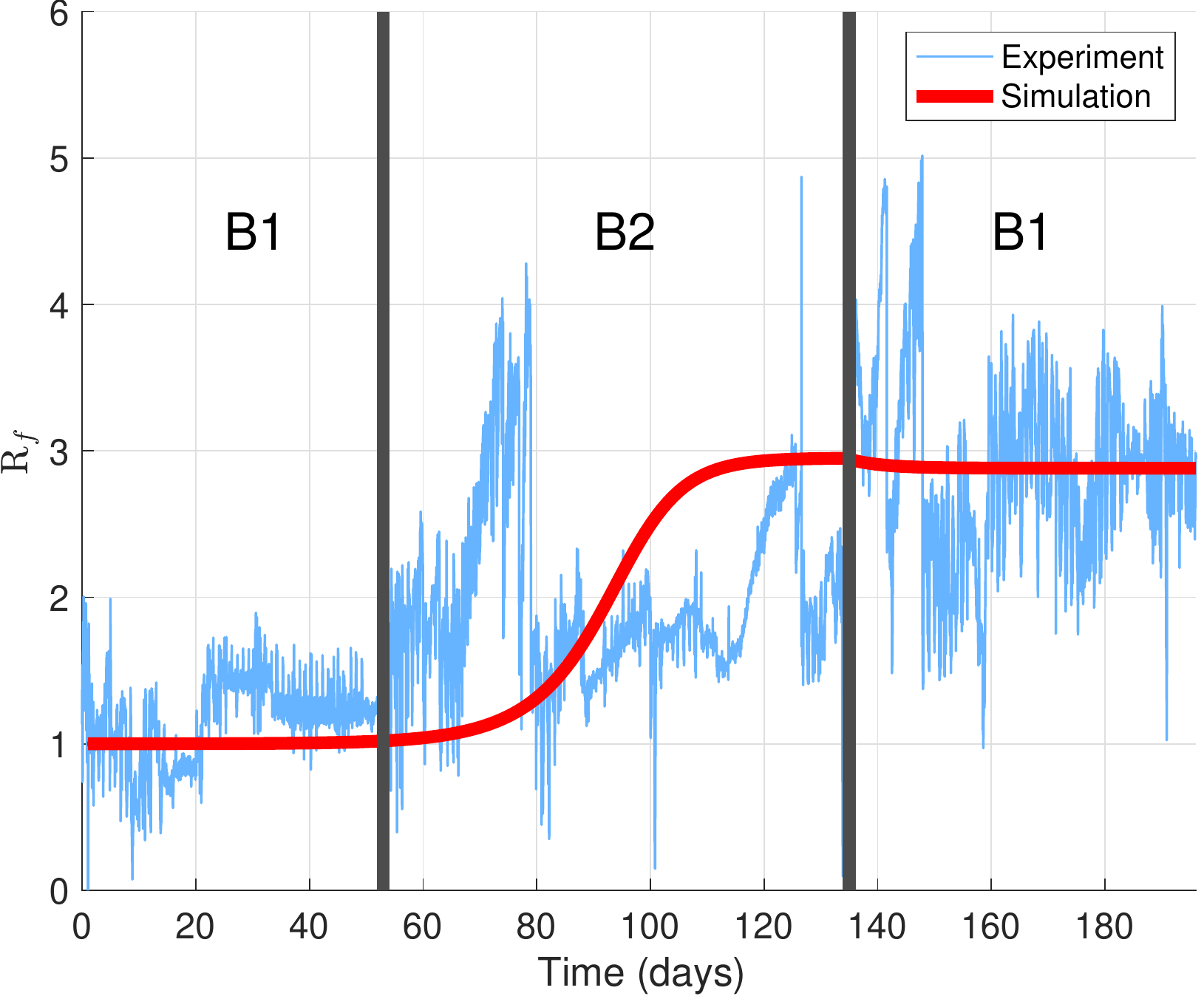}
\caption{---Experimental and simulated resistance factor R$_f$ over time.}
\label{permeability} 
\end{figure}

\captionsetup{justification = raggedright, singlelinecheck = false}
\begin{table}[h!]
\begin{tabular}{p{1.83in} p{4.6in}}		
Name & Equation\\
\hline
Darcy& $v_w=-(k/\mu_w)\partial_x p_w,\hspace{.5cm} \partial_t\phi_f+\partial_xv_w=0$\\
Permeability& $k=k_0(\phi_f/\phi_0)^{\eta}$\\
Porosity& $\phi_f=\phi_0-\phi_b$\\
Substrate& $\partial_t(C_n\phi_f)+\partial_x(v_wC_n-D_n\phi_f\partial_xC_n)=-\mu_n\rho_b\phi_bC_n/(K_n+C_n)$\\
Biofilm&$\partial_t\phi_b=\mu_n\phi_bC_n/(K_n+C_n)-K_{d}\phi_b-K_{str}|\partial_x p_w| \phi_b$\\
\hline
\end{tabular}
\caption{---Core-scale single-phase equations.}
\label{tab:2}
\end{table}
\captionsetup{justification = centering, singlelinecheck = false}
\noindent Nine parameters are needed to solve the mathematical model in Table \ref{tab:2}. To obtain a better estimate of the model parameters, it is necessary to perform various experiments under controlled input quantities. However, these experiments are expensive and time consuming. In \cite{Landa:Article:2019a} a pore-scale mathematical model is calibrated based on the experiments performed by \cite{Liu:Article:2019}, where measurements of the biofilm amount over time are taken for different flux velocities. For the core-scale system described in this section, only one experiment is performed. Then, we select parameter values from the literature in order to run numerical simulations and compare qualitatively the results to the experimental observations. \textbf{Table \ref{parameters}} shows the selected values of these parameters.
\captionsetup{justification = raggedright, singlelinecheck = false}
\begin{table}[h!]
\centering
\begin{tabular}{ p{2.33in} p{1in} p{1.5in} p{1.3in}}		
Parameter & Notation & Value & Reference\\
\hline
Bacterial death rate& $K_{d}$ & 	3.2$\times10^{-6}$ s$^{-1}$&\cite{Kundu:Article:1983}\\ 
Maximum growth rate & $\mu_n$ &   5$\times10^{-6}$ s$^{-1}$&\cite{Li:Article:2011}\\
Monod-half velocity& $K_n$	&	$0.915$ $\text{kg}/\text{m}^3$&\cite{Linville:Article:2013}\\
Substrate diffusion coefficient& $D_n$	& 	5$\times10^{-10}$ $\text{m}^2/\text{s}$ &\cite{Hardy:Article:1993}\\
Biomass density & $\rho_b$	& $20$ $\text{kg}/\text{m}^3$&\cite{Ro:Article:1991}\\
Power law constant & $\eta$	& $2.5$&\cite{Hommel:Article:2018}\\
Water dynamic viscosity & $\mu_w$	& 	$10^{-3}$ Pa$\cdot$s & Standard\\
Water density & $\rho_w$	&	$10^{3}$ $\text{kg}/\text{m}^3$&Standard\\
\hline
\end{tabular}
\caption{---Model parameters for the verification study.}
\label{parameters}
\end{table}
\captionsetup{justification = centering, singlelinecheck = false}
\noindent These parameter values have the same order of magnitude as the ones used in mathematical modeling~\cite{Alpkvist:Article:2007,Duddu:Article:2009,Hommel:Article:2018,Landa:Article:2019a}. The stress coefficient $K_{str}$ is neglected before the simulations and its value is chosen to better fit the experimental measures. In addition to the model parameters and core dimensions, initial, and boundary conditions are needed to complete the model. The initial porosity and permeability are $\phi_0$=0.23 and $k_0$=1528 md respectively. We assume that initially there is no substrate in the brine. The pressure is set to zero initially (in this system, we are only interested in pressure differences that affects the detachment and transport of substrate). Recalling that the time after inoculation and before starting the substrate injection is roughly half day, we assume that the initial volume fraction of biofilm is distributed uniformly along the core and has a value of $\phi_b(\pmb{x},0)=10^{-5}$. Then, we performed numerical simulations using the same injection strategy of substrate. After simulations, the value of the stress coefficient $K_{str}$ that fits best the experimental data is 1.5$\times10^{-9}$ m/Pa$\cdot$s. \textbf{Fig. \ref{permeability}} shows the experimental measurements of the average resistance factor for the core sample and the numerical results over time. 

\noindent From the numerical simulation and experimental data, we observe that the resistance factor changes over time, which means that the permeability is affected by the biofilm. We can distinguish between the three different periods of substrate injection. For the injection of B$_1$, both simulations and experimental measures show that the resistance factor does not increase significantly. When increasing the substrate concentration to B$_2$, we observe that the resistance factor increases due to major biofilm activity. When the substrate injection is set back to B$_1$, we observe a steady state where the resistance factor has increased by a factor of three. From the simulation, it is possible to observe how the resistance factor decreases from changing the substrate injection B$_2$ to B$_1$, which means a sensible response of the biofilm to substrate input. One of the outcomes of this experiment is that the plugging can be controlled by modifying the substrate flux. In addition, the growth conditions of the microorganism are affected by the difference physical environments. From the experimental data we observe fluctuations on the measurements. This behavior is attributed to the dynamical attachment and detachment of biomass, a complex process not included in the mathematical model. However, this mathematical model is simple enough to have few parameters and complex enough to capture processes such as dynamical changes on the resistance factor and a steady state where the biofilm growth is balanced with the detachment and death of bacteria.

\begin{figure}[h!]
\includegraphics[width=\textwidth]{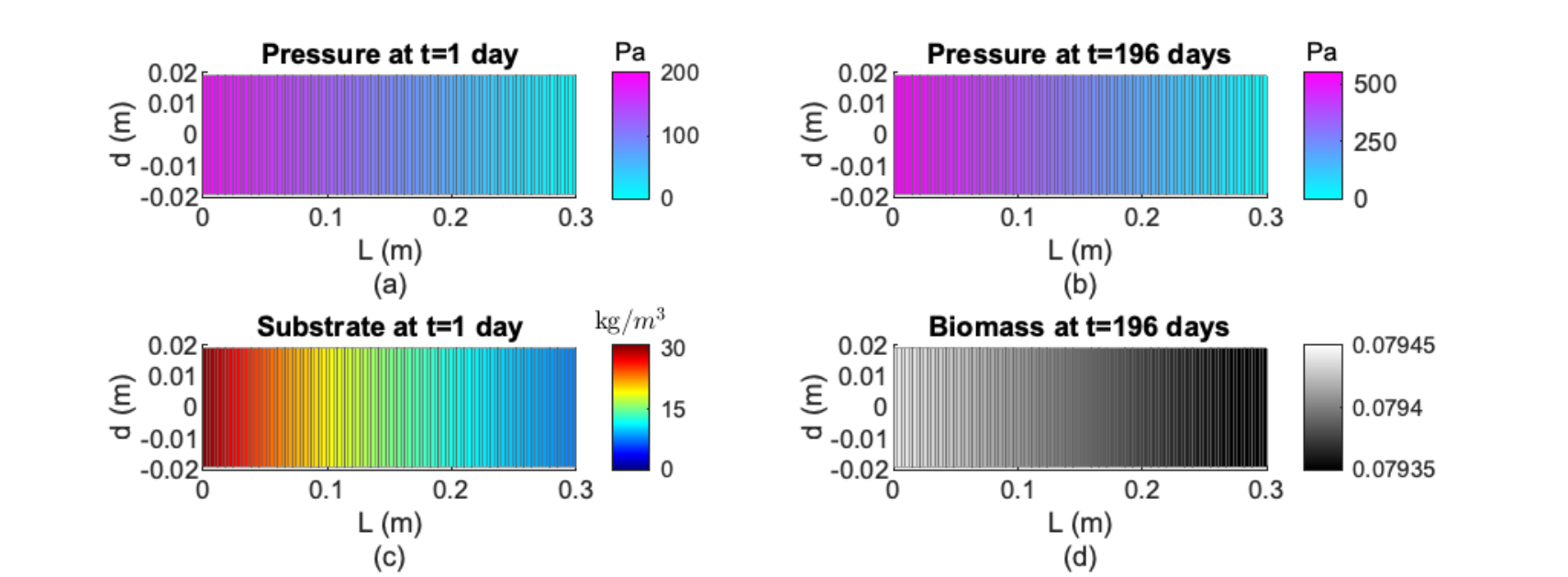}
\caption{---Pressures, substrate, and biomass profiles along the core.}
\label{plotts} 
\end{figure}

\textbf{Fig. \ref{plotts}} shows values of the simulations for pressure, substrate, and biomass along the core. We observe from subplot (a) that after a day of substrate injection the constant flow rate B$_1$ requires an inlet pressure of approximately 200 Pa. From subplot (b) we observe that the inlet pressure has increased more than twice in order to keep the flow rate, as a result of permeability reduction due to the biofilm formation. From the substrate profile (c), this decreases from the inlet to the outlet as bacteria consume them. Given the initial homogeneous volume fraction of biofilm, this has increased more than three order of magnitude in comparison with the initial volume fraction, but it has stopped growing due to the shear forces.   

\section*{Global Sensitivity Analysis}\label{Sensitivity analysis}
\label{sec:glob_sens}
We keep the exposition on sensitivity analysis general to emphasize that the methodology is not restricted to the problems presented in this paper. Let $q(\vec{y})$ be a scalar multidimensional function with $\vec{y} = (y_1,\dots, y_n) \in \Omega \subset \mathbb{R}^{n}$, defined by a range of independent parameters, i.e. $\Omega=(a_1,b_1)\otimes ... \otimes (a_n,b_n)$. We associate a nonnegative weight $w_j(y_j) = 1/(b_j-a_j)$ with each parameter $y_j \in [a_j,b_j]$ for  $j=1,\dots, n$, and use the product measure $w=\prod_{j=1}^{n} w_j$. More general weight functions are possible, including non-product measures corresponding to inter-dependence between the parameters~\cite{Rahman_14}. In this paper we do not aim to quantify uncertainty by stochastic models, but sensitivities in outputs given bounds on the input parameters.

To determine the relative effect of each of the $n$ input parameters on the output quantity of interest $q$, we perform a global sensitivity analysis in terms of a Sobol decomposition~\cite{Sobol_01}. 
The function $q$ is decomposed as a series expansion in all subsets of variables;  the sum of contributions from the individual variables in isolation, all combinations of pairs of variables and so on, leading to the expression:
\begin{equation}
q(\vec{y}) = q^{\{ \emptyset \}} + \sum_{i=1}^{n}q^{\{i\}}(y_i) + \sum_{i=1,j>i}^{n} q^{\{ i,j\}}(y_i,y_j)+\hdots + q^{\{ 1,\hdots, n\}}(\vec{y}),\quad\ldots\ldots\ldots\ldots\ldots\ldots\ldots\ldots\ldots\ldots\ldots\ldots
\end{equation}
where the terms are defined recursively by
\begin{align}
q^{\{ \emptyset \}}  &= \int_{\Omega}q(\vec{y})w(\vec{y})\textup{d}\vec{y},\quad\ldots\ldots\ldots\ldots\ldots\ldots\ldots\ldots\ldots\ldots\ldots\ldots\ldots\ldots\ldots\ldots\ldots\ldots\ldots\ldots\ldots\ldots\ldots\\
q^{\{ i\}}(y_i)  &= \int_{\Omega_{\sim i}}q(\vec{y})w_{\sim i}(\vec{y}_{\sim i}) \textup{d} \vec{y}_{\sim i} - q^{\{ \emptyset \}}, \quad 1\leq i \leq n,\quad\ldots\ldots\ldots\ldots\ldots\ldots\ldots\ldots\ldots\ldots\ldots\ldots\ldots\ldots\\
q^{\{ i, j\}}(y_i,y_j)  &= \int_{\Omega_{\sim i, j}}q(\vec{y})w_{\sim i,j}(\vec{y}_{\sim i,j}) \textup{d} \vec{y}_{\sim i,j} - q^{\{ i\}}(y_i)  - q^{\{ j \}}(y_j) - q^{\{ \emptyset \}}, \quad 1\leq i < j \leq n,\quad\ldots\ldots\ldots\ldots
\end{align}
and so on for higher-order terms. The subscript $\sim i$ means that the $i$th index is omitted, e.g., $w_{\sim i} = \prod_{j \neq i} w_j$. 
The Sobol index for the $s$-parameter combination $\{ y_{i_1},y_{i_2},...,y_{i_s}\}$ is given by
\begin{equation}
S_{ \{ i_1,...,i_s \} } = \frac{1}{\text{Var}(q)}\int_{\Omega_{i_1,...,i_s}} [q^{\{ i_1,...,i_s\}}(y_{i_1},...,y_{i_s})]^2 w_{i_1}(y_{i_1})...w_{i_s}(y_{i_s}) \textup{d}y_{i_1}\hdots \textup{d} y_{i_s}.\quad\ldots\ldots\ldots\ldots\ldots\ldots\ldots
\end{equation}
The total variability in $q$ due to variable $i$ is obtained by summing over all subsets of parameters including parameter $i$, denoted $I_i$, which yields the total Sobol index for parameter $i$,
\begin{equation}
S_{ \{ i \}} = \sum_{s=1}^{n} \sum_{ i \in \{ i_1,...,i_s \} } S_{ \{ i_1,...,i_s \} }.\quad\ldots\ldots\ldots\ldots\ldots\ldots\ldots\ldots\ldots\ldots\ldots\ldots\ldots\ldots\ldots\ldots\ldots\ldots\ldots\ldots\ldots\ldots\ldots
\label{eq:sobol_tot}
\end{equation}

\noindent Directly computing global sensitivities of a computationally complex function, e.g., a computer model, over a domain of only moderately high dimensionality is often infeasible due to the computational cost. As an additional challenge, we are interested in nonsmooth $q$, excluding the use of Sobol decomposition based on generalized polynomial chaos for smooth problems~\cite{Sudret_08}.
As a remedy, we approximate the function of interest with an interpolant on a Clenshaw-Curtis type sparse grid, using the software in~\cite{Klimke_05}. Subsequently, a multi-wavelet decomposition of the interpolant is performed, and the global sensitivity indices are evaluated directly from the multi-wavelet coefficients.
The accuracy in the sensitivity indices as determined by the computer model is thus largely determined by two kinds of errors: interpolation error and basis truncation error introduced when replacing the interpolant by a series expansion in a finite set of multi-wavelets.\\[10pt]
 
\noindent \textbf{Multi-Linear Collocation.} The outline of multi-linear collocation closely follows the exposition in~\cite{Klimke_05}, which also provides the source code for the numerical implementation. We perform interpolation of possibly nonsmooth (but continuous) functions and therefore rely on localized basis functions. For discontinuous functions, we refer to the adaptive sparse grid methods introduced in~\cite{Ma_Zabaras_09}. Multidimensional collocation on sparse grids is built from tensor products of low-order single dimensional collocation rules. Temporarily assume that $n=1$, denote $y=\vec{y}$ and let $Y_{\ell}$ be a set of interpolation nodes in the parameter space, for which the function $q$ is evaluated. The index $\ell$ refers to a refinement level: the higher $\ell$, the bigger the set of interpolation nodes. An interpolant of $q$ is given by
\begin{equation}
\label{eq:interpolant_std}
I_{\ell}(q) = \sum_{y_{\ell} \in Y_{\ell}} q(y_{\ell}) \psi_{y_{\ell}}(y),\quad\ldots\ldots\ldots\ldots\ldots\ldots\ldots\ldots\ldots\ldots\ldots\ldots\ldots\ldots\ldots\ldots\ldots\ldots\ldots\ldots\ldots\ldots\ldots\ldots\ldots
\end{equation}
where $\psi_{y_{\ell}}(y)$ are nodal basis functions with the property
\begin{equation}
\psi_{y_{\ell}}(y_{\ell'}) = \left\{
\begin{array}{ll}
1 &  \text{if } \ell=\ell', \\
0 & \text{otherwise.}\quad\ldots\ldots\ldots\ldots\ldots\ldots\ldots\ldots\ldots\ldots\ldots\ldots\ldots\ldots\ldots\ldots\ldots\ldots\ldots\ldots\ldots\ldots\ldots\ldots
\end{array}
\right.
\end{equation}
In this work we employ the piecewise linear `hat' functions for level $\ell > 1$
\begin{equation}
\psi_{y_{\ell}}(y) = \left\{
\begin{array}{ll}
1 - 2^{i-1} |y-y_{\ell}| &  \text{if } |y-y_{\ell}| < 2^{1-i}, \\
0 &  \text{otherwise, }\quad\ldots\ldots\ldots\ldots\ldots\ldots\ldots\ldots\ldots\ldots\ldots\ldots\ldots\ldots\ldots\ldots\ldots\ldots\ldots
\end{array}
\right.
\end{equation}
supplemented with the unit function (level $\ell =1$). The collocation nodes are nested, i.e., $Y_{\ell} \subset Y_{\ell+1}$. Rather than expressing the interpolant as a function of the nodes in some sufficiently refined set $Y_{\ell}$, it can be expressed as a sum of hierarchical surpluses, i.e., differences between the interpolants at successive levels,
\begin{equation}
\Delta_{\ell} (q) = I_{\ell}(q)-I_{\ell-1}(q), \quad I_{0}(q)\equiv 0.\quad\ldots\ldots\ldots\ldots\ldots\ldots\ldots\ldots\ldots\ldots\ldots\ldots\ldots\ldots\ldots\ldots\ldots\ldots\ldots\ldots\ldots
\end{equation}
Then, Eq.~\eqref{eq:interpolant_std} can be expressed
\begin{equation}
I_{\ell}(q) = \sum_{j=1}^{\ell} \Delta_{j} (q) = \sum_{j=1}^{\ell}  \sum_{y_j \in Y_j \sim Y_{j-1}} q(y_{j}) \psi_{y_j}^{(j)}(Y),\quad\ldots\ldots\ldots\ldots\ldots\ldots\ldots\ldots\ldots\ldots\ldots\ldots\ldots\ldots\ldots\ldots\ldots
\end{equation}
where the superscript $(j)$ is used to emphasize that the basis functions are level specific. Using this construction, one may use the local difference between the interpolant and the true function as a measure to determine where further refinement in terms of more nodes are needed.

In multiple dimensions ($n>1$), a tensor-product formula would lead to very large sets of nodes even for moderate $n$. As a remedy, a sparse grid based on Smolyak's construction~\cite{Smolyak_63} is used, adding only a subset of the tensor product Clenshaw-Curtis nodes at each new level. The hierarchical multidimensional interpolant of $q=q(y_1,\dots, q_n)$ on total level $p\geq n$, is given by
\begin{equation}
\label{eq:md_interp}
I_{p,n}(q) = I_{p-1,n}(q) + \Delta I_{p,n}(q),\quad\ldots\ldots\ldots\ldots\ldots\ldots\ldots\ldots\ldots\ldots\ldots\ldots\ldots\ldots\ldots\ldots\ldots\ldots\ldots\ldots\ldots\ldots\ldots
\end{equation}
with $I_{n-1,n} = 0$ and
\begin{equation}
\Delta I_{p,n}(q) = \sum_{|\vec{i}|=p} (\Delta_{i_1} \otimes \cdots \otimes \Delta_{i_n}) (q) = \sum_{|\vec{i}|=q} \sum_{\vec{j}}  \underbrace{ \left( \psi_{j_1}^{(i_1)} \otimes \cdots \otimes \psi_{j_n}^{(i_n)}  \right)}_{\psi_{\vec{j}}^{(\vec{i})} } \cdot \underbrace{ \left( q(y_{\vec{j}}^{(\vec{i})})  - I_{p-1,n}(q)(y_{\vec{j}}^{(\vec{i})}) \right)}_{w_{\vec{j}}^{(\vec{i})}}.\quad\ldots\ldots
\end{equation}
For more details on the multi-linear collocation method we refer to~\cite{Klimke_05}.
The hierarchical multi-linear interpolant~\eqref{eq:md_interp} is a surrogate from which we can obtain approximations of $q$. The Smolyak algorithm yields the multidimensional interpolant at low computational cost, but the global sensitivity indices are not directly available. Next, we introduce a series expansion of $I_{p,n}$ in multi-wavelets to obtain the Sobol decomposition of $I_{p,n}$.\\[10pt]

\noindent\textbf{Sobol Indices via Multi-Wavelet Expansions.} An alternative representation of $q$ is via a spectral expansion in a set of multidimensional orthogonal basis functions $\{ \varphi_j\}$,
\begin{equation}
q(Y) = \sum_{j=1}^{\infty} c_j \varphi_j(Y),\quad\ldots\ldots\ldots\ldots\ldots\ldots\ldots\ldots\ldots\ldots\ldots\ldots\ldots\ldots\ldots\ldots\ldots\ldots\ldots\ldots\ldots\ldots\ldots\ldots\ldots\ldots
\end{equation}
where the infinite sum is truncated to some finite set in practical application. This is known as generalized polynomial chaos expansion if the basis functions are orthogonal polynomials~\cite{Xiu_Karniadakis_02}, and it has been demonstrated that the Sobol indices can be directly identified from the expansion coefficients, once these have been computed~\cite{Sudret_08}. Nonsmooth functions can be represented by means of multi-resolution analysis where the basis functions are piecewise polynomial multi-wavelets~\cite{LeMaitre_04}, and the identification of the Sobol indices is straightforward.

A piecewise linear multi-wavelet expansion of sufficient resolution can exactly represent the piecewise linear interpolant, provided that the support nodes are aligned with the support of the multi-wavelets. The advantage of using the multi-wavelet expansions rather than trying to directly evaluate the Sobol component functions, is that we may rely on a single interpolation of the function of interest itself (and not its square, conditional on some subset of the variables, and so on).

Let $\{ \varphi_{\vec{k}}(\vec{y}) \}_{\vec{k} \in I_{\text{MW}}}$ be a finite-dimensional basis of multi-wavelets indexed by some set of nonnegative integers $I_{\text{MW}} \subset \mathbb{N}_{0}^{n}$. Ideally, the number of basis functions should be limited but chosen in order to be a good approximation of the interpolant $I_{p,n}(q)$ of the function of interest. The approximation properties of the multi-wavelet basis are determined by the order of the piecewise polynomial multi-wavelets, and the resolution level that governs the localization in parameter space. Analogous to the hierarchical surplus defined for the multi-linear interpolant, the hierarchical surplus of the multi-wavelet expansion is given by the contribution captured by the finest resolution level. The multi-wavelet basis is hierarchical, and can be enriched by adding multi-wavelets to regions labeled important due to large surpluses.

We may now either project the interpolant $I_{p,n}(q)$ onto the basis $\{ \varphi_{\vec{k}}(\vec{y}) \}_{\vec{k} \in I_{\text{MW}}}$, or interpolate the product $q \varphi_{\vec{k}}$, and then perform the projection by computing the expected value. The latter is simpler, as it only involves integration of a piecewise linear function to compute each multi-wavelet coefficient. However, we do not know a priori what multi-wavelets should be included in the truncated basis, and will therefore settle for a simple strategy that also admits adaptivity. Unlike $q$ itself, the interpolant $I_{p,n}(q)$ can be sampled repeatedly at moderate computational cost. Given $N$ random samples of the parameters $\vec{y}$, and a tentative multi-wavelet basis of size $P$, we form the linear system for the $P$-vector of multi-wavelet coefficients $\vec{c}$
\begin{equation}
\label{eq:lin_syst}
\Phi \vec{c} = \vec{q},\quad\ldots\ldots\ldots\ldots\ldots\ldots\ldots\ldots\ldots\ldots\ldots\ldots\ldots\ldots\ldots\ldots\ldots\ldots\ldots\ldots\ldots\ldots\ldots\ldots\ldots\ldots\ldots\ldots\ldots\ldots\ldots
\end{equation}
where the matrix $\Phi \in \mathbb{R}^{N \times P}$ is defined by $[\Phi]_{k,j} = \varphi_{j}(\vec{y}^{(k)})$, and $\vec{q} \in \mathbb{R}^{N}$ contains the evaluations of $I_{p,n}(q)$ in the samples $y^{j}$, $j=1,\dots, N$. The ordinary least squares solution to~\eqref{eq:lin_syst} gives the multi-wavelet coefficients of the tentative bases. If the hierarchical surplus (defined as the difference between the interpolants of two successive grid levels) is above a user specified threshold, the approximation of $I_{p,n}(q)$ is not sufficiently resolved, and we may increase the tentative multi-wavelet basis functions in unresolved regions of parameter space. This is done by solving~\eqref{eq:lin_syst} with an extended basis. As a further measure of how well we represent $I_{p,n}$, we may compare the variance predicted by the multi-wavelet expansion with the sample variance of $I_{p,n}(q)$.

\section*{Numerical Experiments}\label{Numerical experiments}
\begin{figure}[h!]
\includegraphics[width=\textwidth]{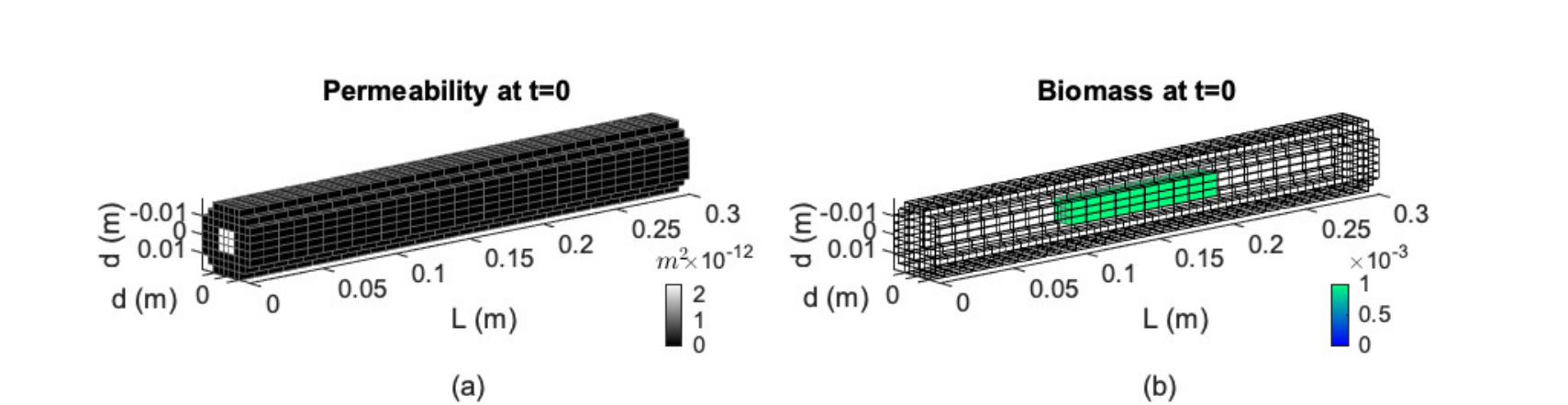}
\caption{---Initial rock permeability (a) and initial volume fraction of biofilm (b).}
\label{permesss} 
\end{figure}

For the numerical experiments, we consider a core-sample saturated with water and oil. We impose a more permeable zone ($10^2k_0$) in the middle of the core, as shown in \textbf{Fig. \ref{permesss}a}. This more permeable zone is set to study the effects of biofilm on oil recovery when it modifies the rock properties in a thief zone. We consider that initially there is only biofilm in the middle part of the thief zone with a volume fraction of 2.5$\times10^{-3}$, as shown in  \mbox{\textbf{Fig. \ref{permesss}b}.}

\noindent The core has an initial water saturation of 0.2 and an oil saturation of 0.8. The substrate is injected on the left side of the core at a velocity of 1 m/day. The initial water pressure is set to 10${}^6$ Pa. \textbf{Table \ref{tab:allparameterss}} lists the parameters for the numerical simulations. These boundary conditions, initial conditions, and parameters are selected to perform the diverse numerical studies in a computational time of order of minutes.
\captionsetup{justification = raggedright, singlelinecheck = false}
\begin{table}[h!]
\centering
\begin{tabular}{p{2.5in} p{1in} p{1.5in} p{1.15in}}		
Parameter & Notation & Values & Reference\\
\hline
Bacterial death rate&$K_{d}$ & 	3.2$\times10^{-6}$ s$^{-1}$&\cite{Kundu:Article:1983}\\ 
Maximum growth rate &$\mu_n$ &    1.59$\times10^{-4}$ s$^{-1}$&\cite{Linville:Article:2013}\\
Monod-half velocity&$K_n$	& 	$0.915$ kg/m$^3$&\cite{Linville:Article:2013}\\
Substrate diffusion&$D_n$	& 	5$\times10^{-10}$ m$^2$/s &\cite{Hardy:Article:1993}\\
Biomass density &$\rho_b$ & $20$ kg/m$^3$&\cite{Ro:Article:1991}\\
Factor Brooks-Corey &$\gamma,\;\beta$ & $2$& \cite{Lacerda:Article:2012}\\
Critical point & $B_c$ & 0.35 & \cite{Vandevivere:Article:1995}\\
Fitting factor Power law & $\eta$ & 2.5& \cite{Hommel:Article:2018} \\
Stress coefficient & $K_{str}$ & $10^{-10}$ m/Pa$\cdot$s & \cite{Landa:Article:2019c}\\
Critical porosity & $\phi_{crit}$ & 0& Assumed\\
Initial permeability & $k_0$ & 2.45$\times10^{-12}$ m$^2$& Assumed\\
Biofilm permeability & $k_b$ & 2.45$\times10^{-13}$ m$^2$ & Assumed\\
Initial porosity & $\phi_0$ & 0.21 & Assumed\\
Core length & $L$ & 0.30 m & Assumed\\
Core radius & $r$ & 0.19 m & Assumed\\
Injected substrate concentration& $C_i$ & 5 kg/m$^3$ & Assumed\\
Injected water velocity & $v_i$ & 1.16$\times10^{-5}$ m/s & Assumed\\
Biofilm water content & $\theta_w$ & 0.9 & Assumed\\
Water viscosity &$\mu_w$	& $10^{-3}$ Pa$\cdot$s & Standard\\
Oil viscosity &$\mu_o$	& 	3.92$\times10^{-3}$ Pa$\cdot$s & Standard\\
Water density &$\rho_w$	& 	$10^{3}$ kg/m$^3$&Standard\\
Oil density &$\rho_o$	& 	$800$ kg/m$^3$&Standard\\
\hline
\end{tabular}
\caption{---Table of input variables and model parameters for the verification study.}
\label{tab:allparameterss}
\end{table}\\[10pt]
\captionsetup{justification = centering, singlelinecheck = false}

\begin{figure}
\includegraphics[width=\textwidth]{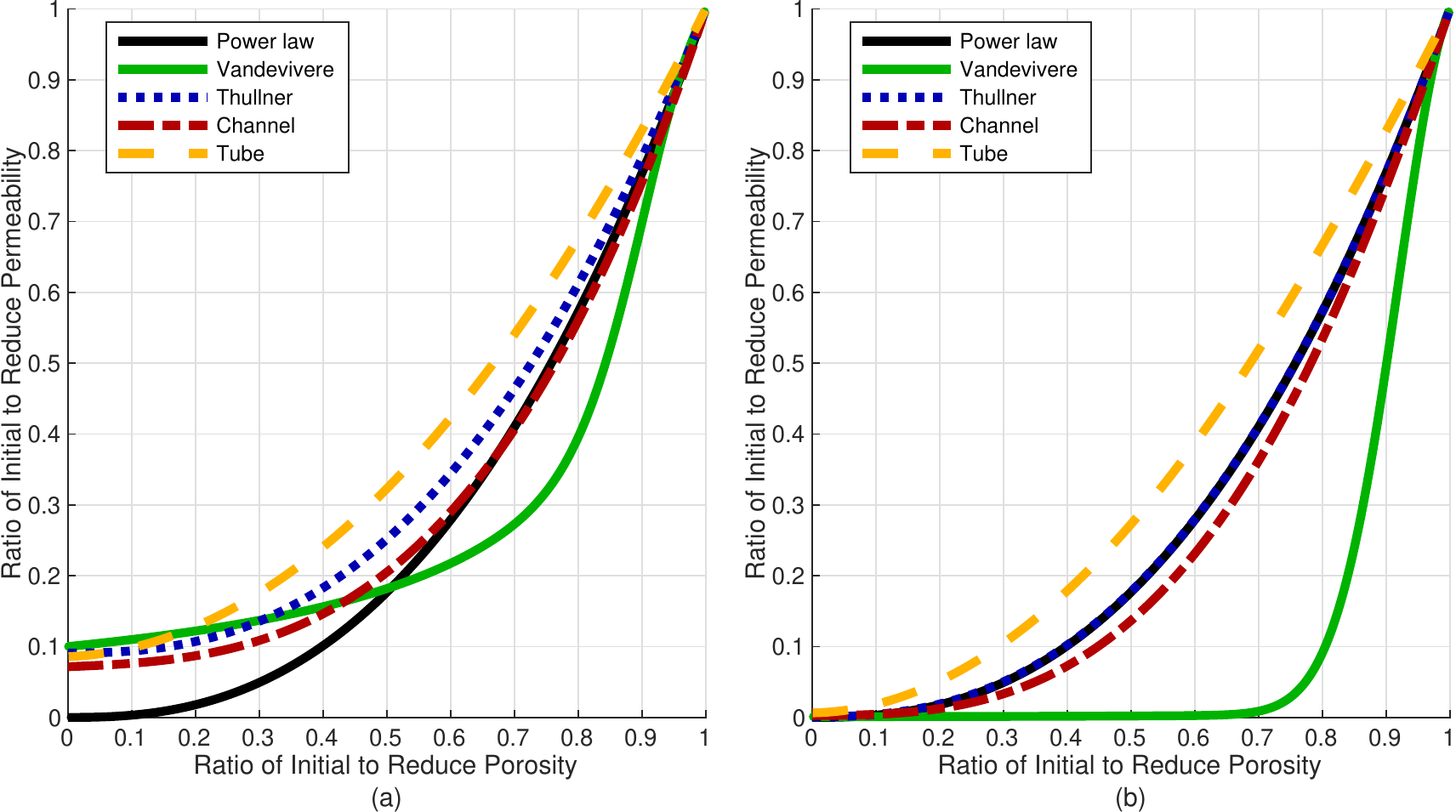}
\caption{---Porosity-permeability relationships of $\pmb{k_b}$=$10^{-1}\pmb{k_0}$ (a) and $\pmb{k_b}$=$10^{-3}\pmb{k_0}$ (b).}
\label{permess} 
\end{figure}

\noindent\textbf{Impact of Porosity-Permeability Relations.} We compare the oil recovery for four porosity-permeability relations: Vandevivere $k_v$ (\ref{Kv}), Thullner $k_{th}$(\ref{Kth}), channel $k_c$ (Appendix A), and  tube $k_t$ (Appendix A). These four relations include the biofilm permeability. \textbf{Fig. \ref{permess}} shows the profiles of these four porosity permeability relations for a highly permeable biofilm and a less permeable biofilm respectively. 

\begin{figure}[h!]
\centering
\includegraphics[width=\textwidth]{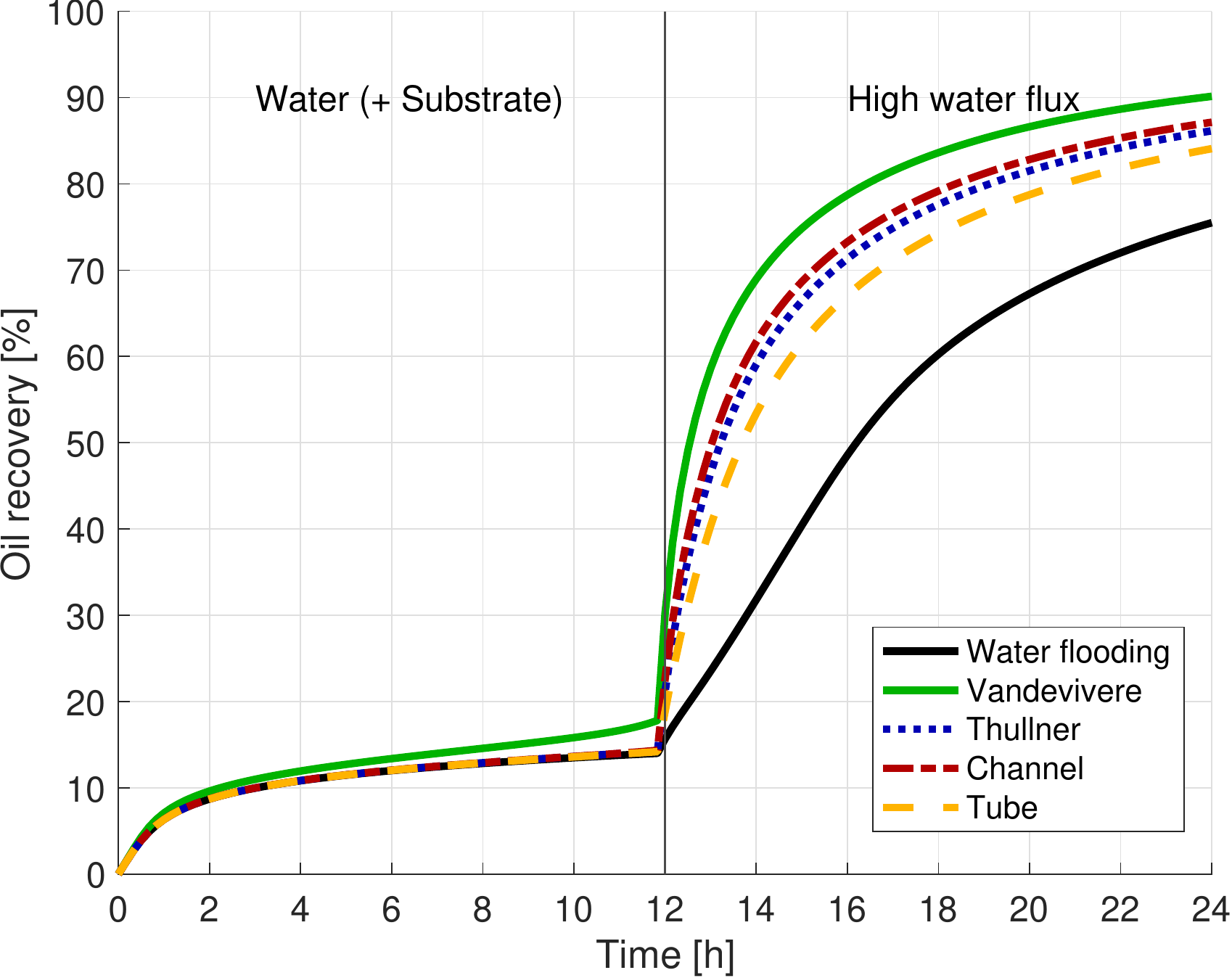}
\caption{---Percentage of oil recoveries for the different porosity-permeability relationships.}
\label{oilrecp} 
\end{figure}

\noindent From Fig. \ref{permess} we can observe that these permeability-porosity relationships give different values of permeability for the same values of rock porosity. Thus, we expect to observe different oil recovery predictions for each of these relationships. \textbf{Fig. \ref{oilrecp}} shows the comparison of percentage of oil recovery in comparison with the initial oil in the core for the four different porosity-permeability relations. The biofilm permeability is set to $k_b=0.1k_0$.

\noindent We observe that after one day of water injection, the Vandevivere porosity-permeability relation predicts larger oil recovery than the tube relation. In addition, the channel relation predicts slightly more oil recovery than the Thullner relation. These results are expected from Fig. \ref{permess}b, where the Vandevivere relationship gives the fastest permeability reduction as a function of the porosity reduction in contrast to the tube relationship which gives the slowest permeability reduction. From Fig.~\ref{oilrecp}, we conclude that the predicted oil recovery differs for the different porosity-permeability curves. The oil recovery could be over- or underestimated depending on the assumed porosity-permeability relation.\\[10pt] 
 
\begin{figure}[h!]
\centering
\includegraphics[width=\textwidth]{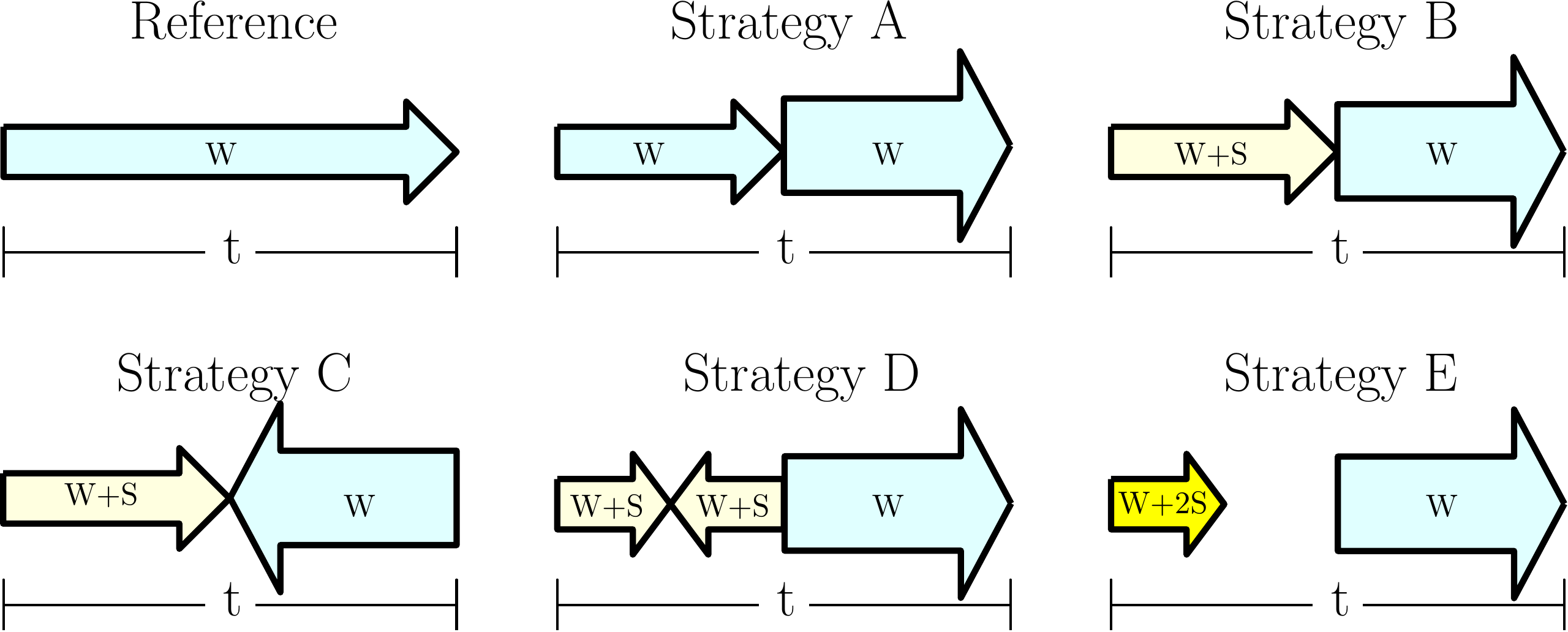}
\caption{---Reference and injection strategies.}
\label{flowdir} 
\end{figure}

\noindent\textbf{Injection Strategies.} The study of injection strategies is important for the optimization of the bioplug technology. Comparison of oil recovery for different injection techniques is possible through numerical experiments. In this work, we compare the oil recovery for different substrate injection strategies for the same initial conditions. For this study, we use the $k_c$ porosity-permeability relation. \textbf{Fig. \ref{flowdir}} shows a reference profile where only water is injected and the five different strategies tested.

\begin{figure}[h!]
\centering
\includegraphics[width=\textwidth]{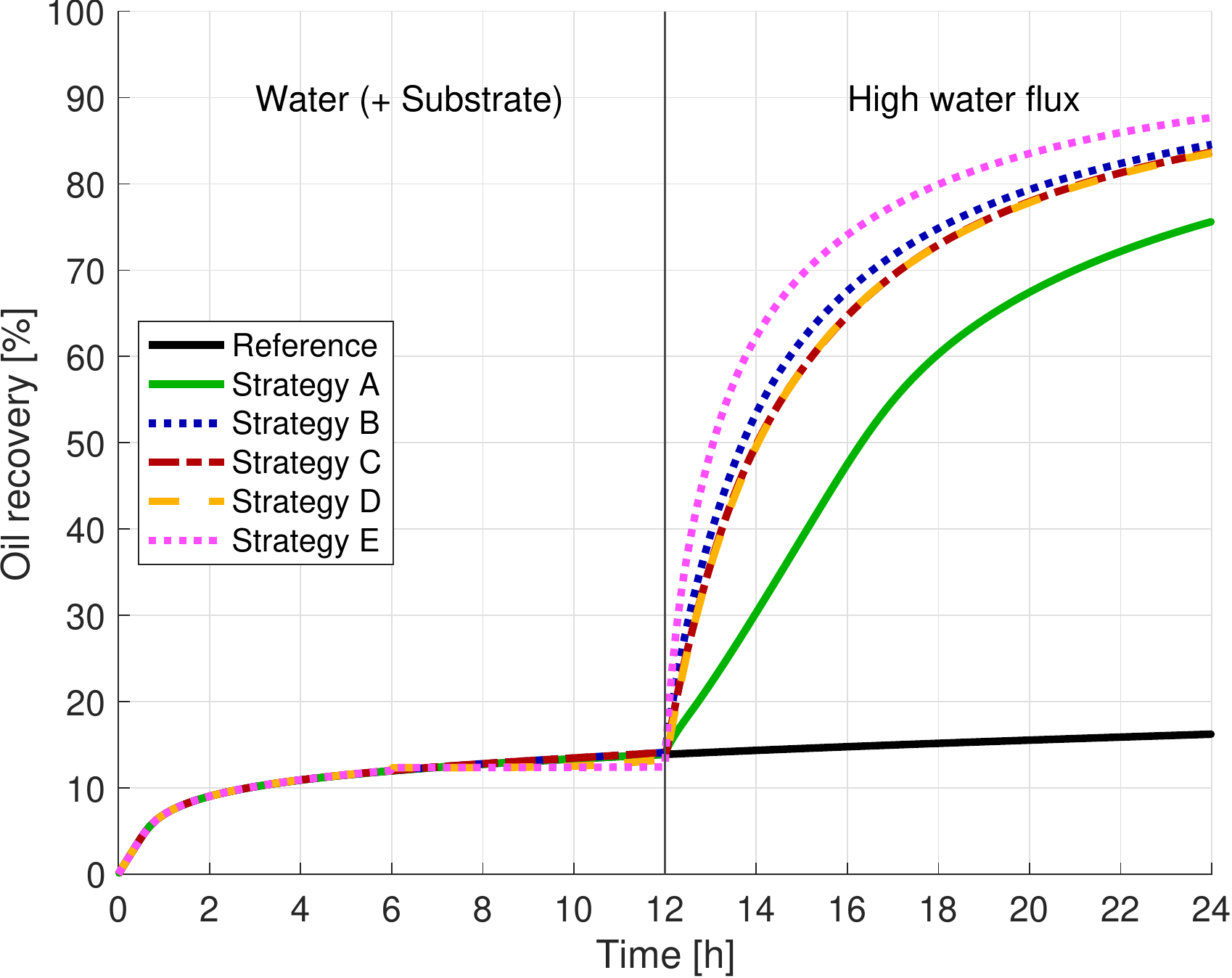}
\caption{---Comparison of the oil recoveries for the different injection strategies over time.}
\label{strategies} 
\end{figure}

The reference strategy consists in injecting only water at a constant flux rate. Strategy A consists in injecting water at a higher flux rate after half the time of the experiment. The second injection strategy is found commonly in the literature. Water and substrate are injected continuously at a fixed flux rate and after increasing the water flow to diverge the water flow. Strategy C consists in inverting the direction of the high flux  to study the effect on the oil recovery in comparison with Strategy B. For Strategy E, first substrate is injected at a fixed flux rate and double concentration. The system is then closed in order to let the bacteria consume the suspended substrate. Afterwards, the water flux is reactivated at the high rate. We observe that Strategy B and Strategy E use the same amount of substrate. \textbf{Fig. \ref{strategies}} shows the three different oil predictions for the different injection strategies.

\begin{figure}[h!]
\includegraphics[width=\textwidth]{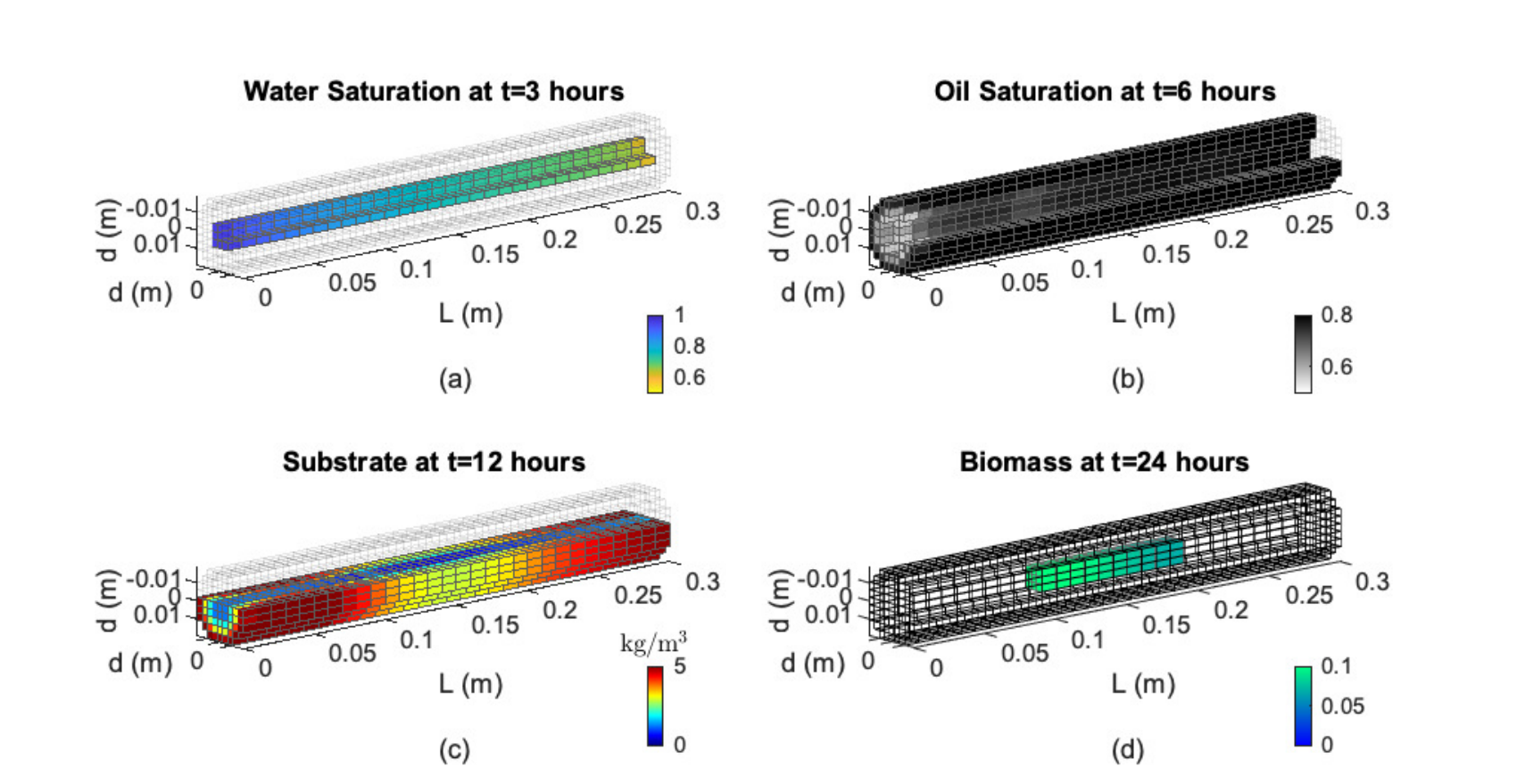}
\caption{---Saturations, substrate, and biomass profiles along the core.}
\label{simulss} 
\end{figure}

From the previous plot we observe that Strategy E predicts the largest oil recovery. Changing the flow direction at half of the injection (Strategy C) does not result in an improvement on the oil recovery in comparison to Strategy B. Strategies C and D give similar predictions of oil recovery. \textbf{Fig. \ref{simulss}} shows the simulation results for pressure, substrate, and biomass along the core for Strategy D.
The cells with water saturation above 0.5 are shown in subplot (a). We observe that after three hours of injection the water has displaced most of the oil in the thief zone. In subplot (b) the cells where oil saturation is above 0.5 after 6 hours of injection are shown. We observe that most of the oil recovered is from the thief zone, but there is still significant amount of oil in the core. Subplot (c) shows the substrate concentration on the lower half part of the core after 12 hours. As described in Strategy D, the flux direction has changed, so now substrate is being injected on the right side of the core, as shown in the figure. From this subplot we observe that the substrate concentration decreases along the core, where the lower values are located in the thief zone. Given the initial homogeneous volume fraction of biofilm, this has increased more than two orders of magnitude in comparison with the initial volume fraction, showing a greater value on the left side. This result shows that changing the substrate flux direction will not lead to a symmetric biofilm formation. 

\noindent\textbf{Sensitivity Analysis: Numerical Results.} Sensitivity analysis is performed for two test cases based on the methodology presented in a previous section.

\textbf{\emph{Case I: Mean Permeability with Variability in Seven Parameters.}} We consider the single-phase core-scale mathematical model in Table \ref{tab:2} and the same boundary and initial conditions as in Model Test. The quantity of interest $q$ is the mean permeability of the core after 100 days of water injection, divided by the initial mean permeability of the core, and we investigate its sensitivity with respect to the seven parameters (i.e., $n=7$) with ranges shown in \textbf{Table~\ref{tab:sens_results_case1}}. The interpolant $I_{p,n}(q)$ is computed for $p=n+0,\dots, n+6$. The relative interpolation error is estimated as the $L_2$ norm of the hierarchical surplus at the finest level over the norm of $I_{p,n}(q)$ itself. The interpolant at level $p=n+6$ requires 30,241 evaluations of $q$, and results in an estimated relative error of less than 0.011. 
\captionsetup{justification = raggedright, singlelinecheck = false}
\begin{table}[h!]
\begin{tabular}{p{2in} p{.75in} p{1.85in} p{1.5in}}
Parameter & Notation & Range & Total Sobol Index\\
\hline
Maximum growth rate & $\mu_{n}$ & [4.5, \ 5.5]$\times10^{-6}$ s$^{-1}$ & 0.76  \\
Bacterial death rate & $K_{d}$ & [2.88, \ 3.52]$\times10^{-6}$ s$^{-1}$ &  0.33 \\
Stress coefficient & $K_{str}$ & [1.35, \ 1.65 ]$\times10^{-9}$ m/Pa$\cdot$s & 0.017  \\
Power law constant & $\eta$ & [2.25, \ 2.75]& 0.0015 \\
Monod-half velocity & $K_n$ & [0.82, \ 1.01] kg/m$^3$ & 0.0017 \\
Biomass density & $\rho_{b}$ & [18, \ 22] kg/m$^3$& 0.0012\\
Nutrient diffusion coefficient & $D_n$ & [4.5, \ 5.5]$\times10^{-10}$ m$^2$/s & 0.0012 \\
\hline
\end{tabular}
\caption{---Total variability contribution of input parameters to percentage of oil extraction.}
\label{tab:sens_results_case1}
\end{table}

\noindent The surrogate function $I_{p,n}(q)$ is subsequently sampled $N=10^5$ times and the multi-wavelet coefficients are computed by solving~\eqref{eq:lin_syst} for varying polynomial orders $o$ and resolution levels $\mathcal{L}$. The fraction of the total sampling variance explained by the retained multi-wavelet coefficients are shown in \textbf{Table~\ref{tab:frac_var_case1}}. For this problem, the convergence is faster in the order $o$ of the piecewise polynomials than in the resolution level $\mathcal{L}$ of the multi-wavelets, suggesting that $q$ is relatively smooth. Only combinations of multi-wavelets of total polynomial order $o$ and total resolution level $\mathcal{L}$ have been included in the bases used to generate the numerical results. The observed accuracy with respect to capturing the total sample variance suggests that this basis truncation is indeed a suitable strategy to reduce the computational cost of solving the linear system~\eqref{eq:lin_syst}.
\begin{table}[h!]
\begin{tabular}{p{.35in}|p{.35in}p{.25in}p{.25in}p{.25in}p{.25in}}
 \diagbox{$\mathcal{L}$}{$o$} & 0 & 1 & 2 & 3 & 4\\ \hline
0 & - & 0.883 & 0.931 & 0.974 & 0.994 \\
1 & 0.695 & 0.901 & 0.949 & 0.980 & 0.997 \\
2 & 0.883 & 0.917 & 0.952 & 0.982 & 0.999 \\
\end{tabular}
\caption{---Fraction of sample variance in test Case I represented by multi-wavelet expansion of polynomial order $o$ and resolution level $\mathcal{L}$.}
\label{tab:frac_var_case1}
\end{table}

\noindent The total contribution of each material parameter, isolated, and in combination with the others, are shown in Table~\ref{tab:sens_results_case1}. Only maximum growth rate and bacterial death rate exhibit significant effect on the variability in $q$ for the parameter ranges considered here.

\textbf{\emph{Case II: Two-Phase Flow with Three Variable Parameters.}} We now consider the case of two-phase flow (oil and water). The injected water velocity changes after 8 hours from $v_i=1$ m/day to a higher velocity of $v_i=50$ m/day. Along the length of the core, we discretize with 30 elements, while on the transversal section we discretize with 9 times 9 elements. Due to the computational complexity, only three parameters are considered in the sensitivity study: initial volume fraction, position, and length of the biofilm, as presented in \textbf{Table~\ref{tab:sens_results_case2}}. The remaining parameters, boundary, and initial conditions are the same as described at the beginning of this section. The quantity of interest is the percentage of oil extraction compared to the initial oil in the core (0-100\%). We expect nonsmooth parameter dependence due to sharp changes in velocity.

\begin{table}[h!]
\begin{tabular}{p{2.75in} p{.6in} p{1.4in} p{1.4in}}
Parameter & Notation & Range & Total Sobol Index\\
\hline
Volume fraction of the biofilm & $\phi_b$ & [2.5$\times10^{-4}$, \ 2.5$\times10^{-3}$] &  $0.86$  \\ 
Length of the biofilm & $L_b$ & [2, \ 20] cm &  $0.18$ \\
Position of the biofilm centre along the core & $X_b$ & [11, \ 19] cm & $0.05$ \\
\hline
\end{tabular}
\caption{---Total contribution of each initial parameter.}
\label{tab:sens_results_case2}
\end{table}

\noindent A sparse grid interpolant on 7 levels yields an estimated error of 0.014 between the two finest levels of resolution, where the relative error has again been estimated as the $L^2$ norm of the hierarchical surplus divided by the norm of the solution itself. A set of $N=10^6$ samples are drawn from the interpolant surrogate model to fit a multi-wavelet model through ordinary least squares. As shown in \textbf{Table~\ref{tab:frac_var_case2}}, essentially all variance is captured by a multi-wavelet expansion in total level 2 and with piecewise quadratic basis functions. The total Sobol indices for this multi-wavelet representation are shown in Table~\ref{tab:sens_results_case2}. For the parameter ranges investigated, the variability in oil extraction is dominated by the initial volume fraction of the biofilm. The length and position of the biofilm should however not be entirely ignored.

\begin{table}[h!]
\begin{tabular}{p{.35in}|p{.4in}p{.4in}p{.4in}p{.4in}}
 \diagbox{$\mathcal{L}$}{$o$} & 0 & 1 & 2 & 3\\ \hline
0 & - & 0.887 & 0.964 & 0.991  \\
1 & 0.662 & 0.901 & 0.970 & 0.991 \\
2 & 0.872 & 0.905 & 0.970 & 0.991  \\
3 & 0.950 & 0.907 & 0.971 & 0.993  \\
\end{tabular}
\caption{---Fraction of sample variance in test Case II represented by multi-wavelet expansion of polynomial order $o$ and resolution level $\mathcal{L}$.}
\label{tab:frac_var_case2}
\end{table}

\section*{Conclusions}\label{Conclusions}
In this work we discuss a core-scale mathematical model for single- and two-phase flow including the transport of substrate and changes on the permeability due to formation of biomass. The single-phase laboratory experiment shows that the substrate input changes the plug-potential. The single-phase mathematical model captured the observed response of permeability to changes in the substrate flux. For the two-phase mathematical model, we investigated the effects on the simulated oil recovery for different empirical and upscaled porosity-permeability relations. These results show that the predicted oil recovery could be over- or underestimated depending on the assumed porosity-permeability relation in the mathematical model. Numerical simulations are performed for different injection strategies to study the oil recovery. After simulations, injecting substrate and stopping the water flow to let the bacteria consume the substrate and after reactivating the flow at a higher rate results in the largest oil recovery prediction. The sensitivity analysis for the single-phase core-scale model shows that two parameters are responsible for almost all variability in the mean permeability: maximum growth rate and bacterial death rate. Both parameters need to be estimated in the laboratory with sufficient accuracy to lead to a reliable estimate of the permeability changes. The sensitivity analysis for the two-phase core-scale model demonstrates less impact on the total variability in oil extraction from the initial position of biofilm as compared to the initial volume fraction and length of the biofilm. Thus, the amount of initial biofilm has a higher impact on the oil recovery in comparison to the initial position of biofilm in the thief zone.\\[10pt] 

\section*{Nomenclature}
$a$ = weighting factor, dimensionless\\
$B_c$, $B_r$ = critical and relative porosity, dimensionless\\
$B_1$, $B_2$= injected brine concentrations, m/L$^3$, kg/m$^3$\\
$C_i$, $C_n$ = injected substrate concentration and substrate concentration, m/L$^3$, kg/m$^3$\\
$d$ = core diameter, L, m\\
$D_n$= substrate diffusion coefficient, L$^2$/s\\
$E$, $F$, $G$, $V$, $W$, $X$ = integration coefficients, dimensionless\\
$g$ = gravity, L/t$^2$, m/s$^2$\\
$\tilde{h}_b$ = biofilm thickness, dimensionless\\
$I_i$ = all subsets of parameters including parameter i for global sensitivity analysis, dimensionless\\
$I_{p,n}(q)$ = hierarchical multidimensional interpolant of q on total level $p\geq n$, dimension dep. on $q$\\
$I_{\text{MW}}$ = multi-wavelet index set of nonnegative integers \\ 
$j_n$ = substrate flux, m/t$\cdot$ L$^2$, kg/s$\cdot$m$^2$\\
$J_\nu$ = Bessel function of order $\nu$ of first kind, dimensionless\\
$k$, $k_0$, $k_b$,  = rock permeability, initial rock permeability, and biofilm permeability, L$^2$, mdarcy [m$^2$]\\
$k_p$, $k_h$, $k_{vp}$ = power law, weighted, and Verma-Pruess permeability relationships, L$^2$, mdarcy [m$^2$]\\
$k_c$, $k_t$, $k_{th}$, $k_v$ = channel, tube, Thullner et al., and Verma-Pruess permeability relationships, L$^2$, mdarcy [m$^2$]\\
$k_{r,o},\;k_{r,w}$ = oil and water relative permeabilities, dimensionless\\
$\tilde{k}_b$ = biofilm permeability, dimensionless\\
$K_{d}$ = bacterial death rate, t$^{-1}$, s$^{-1}$\\
$K_{str}$ = stress coefficient, L$^2\cdot$t/m, m/s$\cdot$Pa\\
$K_n$ = Monod half-velocity coefficient, m/L$^3$, kg/m$^3$\\
$\ell$ = refinement level, dimensionless\\
$\mathcal{L}$ = number of resolution levels, dimensionless\\
$L$, $L_b$ = core and biofilm length, L, cm\\
$n$ = number of stochastic dimensions, dimensionless\\
$\mathbb{N}_{0}^{n}$ = set of $n$-tuples of nonnegative integers, dimensionless\\
$o$ = polynomial order of the multi-wavelet expansion, dimensionless\\
$p_o$, $p_w$ = oil and water pressure, m/L$\cdot$t$^2$, Pa\\
$P$ = size of multi-wavelet basis\\
$q$ = quantity of interest for global sensitivity analysis, varying dimension\\
$r$ = core radius, L, cm\\
$R_f$ = resistance factor, dimensionless\\
$R_n$ = substrate reaction term, m/t$\cdot$L$^3$, kg/s$\cdot$m$^3$\\
$S_o,\;S_w$ = saturation of oil and water, dimensionless\\
$S_{\lbrace i\rbrace}$ = total Sobol index for parameter $i$, dimensionless\\
$t$ = time, t, days [hours]\\
$T$ = temperature, T, \textdegree{}C\\  
$v_i$, $v_o$, $v_w$ = injected water velocity, oil, and water velocity, L/t, m/s\\
$w$ = variable depending on the biofilm thickness, dimensionless\\
$X_b$ = position of the biofilm centre along the core, L, cm\\
$y_j$ = general parameter for global sensitivity analysis, dimensionless\\
$Y$ = interpolation nodes in the parameter space\\
$Y_\nu$ = Bessel function of order $\nu$ of second kind, dimensionless\\
$Y_{\ell}$ = set of interpolation nodes in the parameter space, dimensionless\\
$\beta$ = fitting factor Brooks-Corey relationship, dimensionless\\
$\gamma$ = factor Brooks-Corey relationship, dimensionless\\
$\eta$ = fitting factor Power law, dimensionless\\
$\theta_w$ = biofilm water content, dimensionless\\
$\mu_n$ = maximum specific biomass production rate, t$^{-1}$, s$^{-1}$\\
$\mu_o$, $\mu_w$ = water and oil viscosity, m/L$\cdot$t, Pa$\cdot$s\\
$\xi$ = variable dependent on biofilm permeability and porosity, dimensionless\\
$\rho_b,\;\rho_o,\;\rho_w$ = density of biomass, oil, and water, m/L$^3$, kg/m$^3$\\
$\phi$ = rock porosity, dimensionless\\
$\phi_b$, $\phi_f$ = volume fraction of biofilm and void space outside the biofilm, dimensionless\\
$\varphi_j$ = $j$th orthogonal basis function, dimensionless\\ 
$\psi_{y_{\ell}}$ = piecewise linear interpolation functions, dimensionless\\
$\Omega$ = range of independent parameters for global sensitivity analysis, dimensionless\\[10pt]

\noindent\textbf{Acknowledgments} The work of DLM, GB, BFV, KK, PP, and FAR was partially supported by the Research Council of Norway through the projects IMMENS no. 255426, MICAP no. 268390, and CHI no. 255510. ISP was supported by the Research Foundation-Flanders (FWO), Belgium through the Odysseus program (project G0G1316N) and the Akademia grant of Equinor ASA. The authors also appreciate the support from Equinor ASA related to the experimental work reported herein.

\bibliographystyle{numeric}

\begin{thebibliography}{}
\expandafter\ifx\csname natexlab\endcsname\relax\def\natexlab#1{#1}\fi
\def\au#1{#1} \def\ed#1{#1} \def\yr#1{#1}\def\at#1{#1}\def\jt#1{\textit{#1}}
  \def\bt#1{#1}\def\bvol#1{\textbf{#1}} \def\vol#1{#1} \def\pg#1{#1}
  \def\publ#1{#1}\def\arxiv#1{#1}\def\org#1{#1}\def\st#1{\textit{#1}}
  
\bibitem{Alpkvist:Article:2007}
{\sc \au{Alpkvist, E.} \& \au{Klapper, I.}} \yr{2007}  \at{A multidimensional
  multispecies continuum model for heterogeneous biofilm development}.
  \jt{Bull.\ Math.\ Biol.}  \bvol{69}~(2),  \pg{765--789}.
  
\bibitem{Bao:Article:2017}
{\sc \au{Bao, K.}, \au{Lie, K.-A.}, \au{M{\o}yner, O.} \& \au{Liu, M.}}
\yr{2017} \at{Fully implicit simulation of polymer flooding with MRST}.
  \jt{Comput.\ Geosci.}  \bvol{21}~(5-6),  \pg{1219--1244}.
  
\bibitem{Brockmann:Article:2006}
{\sc \au{Brockmann, D.}, \au{Rosenwinkel, K.-H.} \& \au{Morgenroth, E.}}
\yr{2006} \at{Modelling deammonification in biofilm systems: Sensitivity and identifiability analysis as a basis for the design of experiments for parameter estimation}. \jt{Comput. Aided Chem. Eng.}  \bvol{21}, \pg{221--226}.

\bibitem{Carman:Article:1937}
{\sc \au{Carman, P.C.}} \yr{1937} \at{Fluid flow through granular beds}.
\jt{Trans. Inst. Chem. Eng.} \bvol{15}, \pg{150166}.
  
\bibitem{Corey:Article:1954}
{\sc \au{Corey, A.T.}} \yr{1954} \at{The interrelation between gas and oil relative permeabilities}.
\jt{Prod. Mon.} \bvol{19}, \pg{38--42}. 
   
\bibitem{Duddu:Article:2009}
{\sc \au{Duddu, R.}, \au{Chopp, D.~L.} \& \au{Moran, B.}} \yr{2009}  \at{A
  two-dimensional continuum model of biofilm growth incorporating fluid flow
  and shear stress based detachment}.  \jt{Biotechnol.\ Bioeng.}
  \bvol{103}~(1),  \pg{92--104}.
  
\bibitem{Hardy:Article:1993}
{\sc \au{Hardy, B., Sarko, A.}} \yr{1993} \at{Molecular dynamics simulation of cellobiose in water}.
\jt{J. Comput. Chem.}  \bvol{14}~(7),  \pg{848--857}.
  
\bibitem{Hommel:Article:2018}
{\sc \au{Hommel, J.}, \au{Coltman, E.} \& \au{Class, H.}} \yr{2018}
  \at{Porosity--permeability relations for evolving pore space: A review with a
  focus on (bio-)geochemically altered porous media}.  \jt{Transp.\ Porous
  Med.}  \bvol{124}~(2),  \pg{589--629}.
  
\bibitem{Ives:Article:1965}
{\sc \au{Ives, K.} \& \au{Pienvichitr, V.}} \yr{1965}
\at{Kinetics of the filtration of dilute suspensions}.
\jt{Chem. Eng. Sci.}  \bvol{20}~(11),  \pg{965--973}.

\bibitem{Kim:Article:2006}
{\sc \au{Kim, S.B.}} \yr{2006}
\at{Numerical analysis of bacteria transport in saturated porous media}.
\jt{Hydrol. Process.}  \bvol{20}~(5),  \pg{1177--1186}.

\bibitem{Klimke_05}
{\sc Klimke, A., \& Wohlmuth, B.~I.} \yr{2005}
\at{ Algorithm 847: Spinterp: piecewise multilinear hierarchical sparse
  grid interpolation in {M}atlab.}
\jt{ACM Trans. Math. Softw.} \bvol{31} \pg{561--579}.

\bibitem{Kundu:Article:1983}
{\sc \au{Kundu, S.}, \au{Ghose, T.K.} \& \au{Mukhopadhyay, S.N.}} \yr{1983}
\at{Bioconversion of cellulose into ethanol by \textit{Clostridium thermocellum} -- product inhibition}.
\jt{Biotechnol. Bioeng.}  \bvol{25}~(4),  \pg{1109--1126}.

\bibitem{Lacerda:Article:2012}
{\sc \au{Lacerda, E.C.D.S.}, \au{Priimenko, V.I.} \& \au{Pires, A.P.}} \yr{2012}
\at{Microbial EOR: A Quantitative Prediction of Recovery Factor}.
\jt{Society of Petroleum Engineers}.

\bibitem{Landa:Article:2019a}
{\sc \au{Landa-Marb\'an, D.}, \au{Liu, N.}, \au{Pop, I.~S.}, \au{Kumar, K.},
  \au{Pettersson, P.}, \au{B{\o}dtker, G.}, \au{Skauge, T.} \& \au{Radu,
  F.~A.}} \yr{2019}  \at{A pore-scale model for permeable biofilm: Numerical
  simulations and laboratory experiments}.  \jt{Transp. Porous Med.}
  \bvol{127}~(3), \pg{643--660}.

\bibitem{Landa:Article:2019}
{\sc \au{Landa-Marb\'an, D.}, \au{B{\o}dtker, G.}, \au{Kumar, K.}, \au{Pop, I.~S.},  \& \au{Radu,
  F.~A.}} \at{An upscaled model for permeable biofilm in a thin channel and tube}. \bvol{Submitted}
  
\bibitem{Landa:Article:2019c}
{\sc \au{Landa-Marb\'an, D.}, \au{Pop, I.S.}, \au{Kumar, K.} \& \au{Radu, F.A.}}
\yr{2019} \at{Numerical Simulation of Biofilm Formation in a Microchannel}.
\bt{In {\em Numerical Mathematics and Advanced Applications ENUMATH 2017\/} (ed. \ed{Radu, F.A., Kumar, K., Berre, I., Nordbotten, J.M. \& Pop, I.S.r})}, \bvol{126}, \pg{799--807}, \publ{Springer International Publishing, Cham}.

\bibitem{LeMaitre_04}
{\sc \au{Le~Ma\^{i}tre, O.}, \au{Najm, H.}, \au{Ghanem, R.}, \& \au{Knio, O.}} \yr{2004}
\at{Multi-resolution analysis of {W}iener-type uncertainty propagation
  schemes.}
\jt{J. Comput. Phys. 197}, \bvol{2} \pg{502--531}.
    
\bibitem{Li:Article:2011}
{\sc \au{Li, J.}, \au{Liu, J.}, \au{Trefry, M.~G.}, \au{Park, J.}, \au{Liu, K.}, \au{Haq, B.}, \au{Johnston, C.~D.},
\& \au{Volk, H.}} \yr{2011} \at{Interactions of Microbial-Enhanced Oil Recovery Processes.}
 \jt{Transp. Porous Med.} 
\bvol{87}~(1), \pg{77--104}.

\bibitem{Lie:Book:2019}
{\sc \au{Lie, K.-A.}} \yr{2019}  \at{An Introduction to Reservoir Simulation Using MATLAB/GNU Octave: User Guide to the MATLAB Reservoir Simulation Toolbox (MRST)}. \publ{Cambridge University Press}.

\bibitem{Linville:Article:2013}
{\sc \au{Linville, J.}, \au{Rodriguez, M.}, \au{Mielenz, J.} \& \au{Cox, C.}} \yr{2013}
\at{Kinetic modeling of batch fermentation for \textit{Populus} hydrolysate tolerant mutant and wild type strains of \textit{Clostridium Thermocellum}}. \jt{Bioresour. Technol.}  \bvol{147},  \pg{605--613}.

\bibitem{Liu:Article:2019}
{\sc \au{Liu, N.}, \au{Skauge, T.}, \au{Landa-Marb\'an, D.}, \au{Hovland, B.},
  \au{Thorbj{\o}rnsen, B.}, \au{Radu, F.~A.}, \au{Vik, B.~F.}, \au{Baumann, T.}
  \& \au{B{\o}dtker, G.}} \yr{2019}  \at{Microfluidic study of effects of
  flowrate and nutrient concentration on biofilm accumulation and adhesive
  strength in a microchannel}.  \jt{J. Ind. Microbiol. Biotechnol.}
  \bvol{46}~(6),  \pg{855--868}.
  
  \bibitem{Ma_Zabaras_09}
{\sc \au{Ma, X.} \& \au{Zabaras, N.}} \yr{2009}
\at{An adaptive hierarchical sparse grid collocation algorithm for the
  solution of stochastic differential equations.}
\jt{J. Comput. Phys.} \bvol{228}  \pg{3084--3113}.
  
\bibitem{Patel:Article:2015}
{\sc \au{Patel, I.}, \au{Borgohain, S.}, \au{Kumar, M.}, \au{Rangarajan, V.}, \au{Somasundaran, P.} \& \au{Sen, R.}} \yr{2015} \at{Recent developments in microbial enhanced oil recovery}. \jt{Renew. Sustain. Energy Rev.}  \bvol{52} \pg{1539--1558}.

\bibitem{Rahman_14}
{\sc \au{Rahman, S.}} \yr{2014}
\at{A generalized {ANOVA} dimensional decomposition for dependent
  probability measures.}
\jt{SIAM/ASA J. Uncertain. Quantif.} \bvol{2} \pg{670--697}.

\bibitem{Ro:Article:1991}
{\sc \au{Ro, K. S.} \& \au{Neethling, J. B.}} \yr{1991}
\at{Biofilm density for biological fluidized beds}.
\jt{Res. J. Water Pollut. Control Fed.}
\bvol{63}, \pg{815--818}
  
\bibitem{Schulz:Article:2017}
{\sc \au{Schulz, R.} \& \au{Knabner, P.}} \yr{2016}  \at{Derivation and
  analysis of an effective model for biofilm growth in evolving porous media}.
  \jt{Math.\ Methods Appl.\ Sci.}  \bvol{40}~(8),  \pg{2930--2948}.
  
  \bibitem{Smolyak_63}
{\sc \au{Smolyak, S.}} \yr{1963}
\at{Quadrature and interpolation formulas for tensor products of certain
  classes of functions.}
\jt{Soviet Mathematics, Doklady} \bvol{4} \pg{240--243}.

\bibitem{Sobol_01}
{\sc Sobol, I.}
\newblock Global sensitivity indices for nonlinear mathematical models and
  their {M}onte {C}arlo estimates.
\jt{Math.Comput. Simulat.} \bvol{55}~(1) \yr{2001} \pg{271--280}.
  
  \bibitem{Sudret_08}
{\sc \au{Sudret, B.}} \yr{2008}
\at{Global sensitivity analysis using polynomial chaos expansions.}
\jt{Reliab. Eng. Syst. Safe.} \bvol{93}~(7) \pg{964--979}.

\bibitem{Suthar:Article:2009}
{\sc \au{Suthar, H.}, \au{Hingurao, K.}, \au{Desai, A.} \& \au{Nerurkar, A.}} \yr{2009}
\at{Selective Plugging Strategy-Based Microbial-Enhanced Oil Recovery Using \textit{Bacillus licheniformis} TT33}.
\jt{J. Microbiol. Biotechnol.}
 \bvol{19}~(10),  \pg{1230--1237}.

\bibitem{Thullner:Article:2002}
{\sc \au{Thullner, M.}, \au{Zeyer, J.} \& \au{Kinzelbach, W.}} \yr{2002}
  \at{Influence of microbial growth on hydraulic properties of pore networks}.
  \jt{Transp.\ Porous Med.}  \bvol{49}~(1),  \pg{99122}.
  
\bibitem{vanNoorden:Article:2010}
{\sc \au{{van Noorden}, T.~L.}, \au{Pop, I.~S.}, \au{Ebigbo, A.} \& \au{Helmig,
  R.}} \yr{2010}  \at{An upscaled model for biofilm growth in a thin strip}.
  \jt{Water Resour.\ Res.}  \bvol{46}~(6),  \pg{W06505}. 
  
\bibitem{Vandevivere:Article:1995}
{\sc \au{Vandevivere, P.}} \yr{1995}  \at{Bacterial clogging of porous media: a
  new modelling approach}.  \jt{Biofouling}  \bvol{8}~(4),  \pg{281291}. 
  
\bibitem{Verma:Article:1988}
{\sc \au{Verma, A.} \& \au{Pruess, K.}} \yr{1988}
\at{Thermohydrological conditions and silica redistribution near high-level nuclear wastes emplaced in saturated geological formations}. \jt{J. Geophys. Res.: Solid Earth} \bvol{93}~(B$_2$),  \pg{1159--1173}. 

\bibitem{Wood:Article:2019}
{\sc \au{Wood, D.A.}} \yr{2019}
\at{Microbial improved and enhanced oil recovery (MIEOR): Review of a set of technologies diversifying their applications}. \jt{Advances in Geo-EnergyResearch} \bvol{3}~(2),  \pg{122--140}. 

\bibitem{Xiu_Karniadakis_02}
{\sc \au{Xiu, D.} \& \au{Karniadakis, G.~E.}} \yr{2002}
\at{The {W}iener--{A}skey polynomial chaos for stochastic differential
  equations}
\jt{SIAM J. Sci. Comput. 24}, \bvol{2} \pg{619--644}.

\end{thebibliography}

\setcounter{equation}{0}
\renewcommand{\theequation}{A-\arabic{equation}}
\section*{Appendix A --- Effective Porosity-Permeability Relations}\label{appendix}
A detailed description of the following two porosity-permeability relationships can be found in \cite{Landa:Article:2019}, where both relationships are derived by homogenization of a pore-scale model. To this aim we let $\tilde{h}_b$ be the dimensionless thickness of the biofilm layer, $\phi_b$ the volume fraction of biofilm, $\phi_0$ the initial porosity, $\theta_w$ the biofilm porosity, and $k_b$ the biofilm permeability. The thickness of the biofilm $\tilde{h}_b$ is given as a function of the volume fraction of biofilm $\phi_b$ and the initial porosity $\phi_0$ for the thin channels as $\tilde{h}_b=\phi_b/\phi_0$, whereas for the thin tubes $\tilde{h}_b=1-\sqrt{1-\phi_b/\phi_0}$. We use the notation $w=1-\tilde{h}_b$ and $\tilde{k}_b=k_b/k_0$.

The effective porosity-permeability relation for a porous medium modeled as a stack of thin channels is given by
\begin{equation}
\frac{k_c}{k_0}=-\frac{w^3}{6}-wV-\frac{W\exp\left(-\lambda\right) \left[\exp(\tilde{h}_b\lambda)-1\right]-X\exp(\lambda) \left[\exp(-\tilde{h}_b\lambda)-1\right]}{\lambda}+{\tilde{k}_b}\tilde{h}_b,\quad\ldots\ldots\ldots\ldots\ldots\ldots
\end{equation}
where
\begin{eqnarray}
V=&&\frac{\left(w^2+2\tilde{k}_b\right)\left[\exp\left(-\tilde{h}_b\lambda\right)+\exp\left(\tilde{h}_b\lambda\right)\right]+2w\sqrt{\tilde{k}_b\theta_w}
    \left[\exp\left(\tilde{h}_b\lambda\right)-\exp\left(-\tilde{h}_b\lambda\right)\right]-4{\tilde{k}_b}}{2\left[\exp\left(-\tilde{h}_b\lambda\right)+\exp\left(\tilde{h}_b\lambda\right)\right]},\quad\ldots\\
W=&&\frac{{\tilde{k}_b}\exp\left(w\lambda\right)-w\sqrt{\tilde{k}_b\theta_w}\exp\left(\lambda\right)}{\exp\left(-\tilde{h}_b\lambda\right)+\exp\left(\tilde{h}_b\lambda\right)},\quad\ldots\ldots\ldots\ldots\ldots\ldots\ldots\ldots\ldots\ldots\ldots\ldots\ldots\ldots\ldots\ldots\ldots\ldots\ldots\\
X=&&\frac{{\tilde{k}_b}\exp\left(-w\lambda\right)+w\sqrt{\tilde{k}_b\theta_w}\exp\left(-\lambda\right)}{\exp\left(-\tilde{h}_b\lambda\right)+\exp\left(\tilde{h}_b\lambda\right)},\quad\ldots\ldots\ldots\ldots\ldots\ldots\ldots\ldots\ldots\ldots\ldots\ldots\ldots\ldots\ldots\ldots\ldots\ldots
\end{eqnarray} 
where $\lambda=\sqrt{\theta_w/{\tilde{k}_b}}$.

The effective porosity-permeability relation for a porous medium modeled as a stack of thin tubes is given by
\begin{eqnarray}
\frac{k_t}{k_0}=-\frac{w^4}{8}-w^2E+\frac{2\left(Y_1\left(-\xi\right)F-J_1\left(\xi\right)G-wY_1\left(-w\xi\right)F+wJ_1\left(w\xi\right)G\right)}{\xi}+\tilde{k}_b\left(1-w^2\right)\quad\ldots\ldots\ldots\ldots
\end{eqnarray}
where
\begin{eqnarray}
E=&&\frac{2w\theta_w\left[J_0\left(\xi\right)Y_0\left(-w\xi\right)-J_0\left(w\xi\right)Y_0\left(-\xi\right)\right]+\xi {\tilde{k}_b}\left[J_0\left(w\xi\right)Y_1\left(-w\xi\right)+Y_0\left(-w\xi\right)J_1\left(w\xi\right)\right]}{4\left[\xi J_0\left(\xi\right)Y_1\left(-w\xi\right)+\xi Y_0\left(-\xi\right)J_1\left(w\xi\right)\right]}\nonumber\\
&&-\frac{\xi\left(4{\tilde{k}_b}+w^2\right)\left[J_0\left(\xi\right)Y_1\left(-w\xi\right)+Y_0\left(-\xi\right)J_1\left(w\xi\right)\right]}{4\left[\xi J_0\left(\xi\right)Y_1\left(-w\xi\right)+\xi Y_0\left(-\xi\right)J_1\left(w\xi\right)\right]},\quad\ldots\ldots\ldots\ldots\ldots\ldots\ldots\ldots\ldots\ldots\ldots\ldots\ldots\\
F=&&\frac{2{\tilde{k}_b}\xi Y_1\left(-w\xi\right)+w\theta_wY_0\left(-\xi\right)}{2\left[\xi J_0\left(\xi\right)Y_1\left(-w\xi\right)+\xi Y_0\left(-\xi\right)J_1\left(w\xi\right)\right]},\quad\ldots\ldots\ldots\ldots\ldots\ldots\ldots\ldots\ldots\ldots\ldots\ldots\ldots\ldots\ldots\ldots\ldots\\
G=&&\frac{2{\tilde{k}_b}\xi J_1\left(w\xi\right)+w\theta_wJ_0\left(\xi\right)}{2\left[\xi J_0\left(\xi\right)Y_1\left(-w\xi\right)+\xi Y_0\left(-\xi\right)J_1\left(w\xi\right)\right]},\quad\ldots\ldots\ldots\ldots\ldots\ldots\ldots\ldots\ldots\ldots\ldots\ldots\ldots\ldots\ldots\ldots\ldots
\end{eqnarray}
where $\xi = i\sqrt{\theta_w/{\tilde{k}_b}}$ and $i$ is the imaginary number. Here, $J_\nu\left({{z}}\right)$ and $Y_\nu\left({{z}}\right)$ are the Bessel function of order $\nu$ of first and second kind respectively.\\[15pt]

\noindent\textbf{David Landa-Marb\'an} is a post-doctor at the Norwegian Research Centre at the Energy Department. He is interested in multiscale modeling for biofilms. Landa-Marb\'an holds a PhD degree in applied mathematics from the University of Bergen.\\[8pt]
\noindent\textbf{Gunhild B{\o}dtker} is a senior researcher at the Norwegian Research Centre, NORCE. B{\o}dtker is research director for the group integrated microbiology, chemistry, and physics at the NORCE Energy Department. She is interested in reservoir microbiology, MEOR, reservoir souring, and biofilm injectivity. B{\o}dtker holds a PhD degree in microbiology from the University of Bergen.\\[8pt]
\noindent\textbf{Bartek Florczyk Vik} is a senior researcher at the Norwegian Research Centre at the Energy Department. He is interested in laboratory experiments for petroleum microbiology. Vik holds a PhD degree in physics from the University of Bergen.\\[8pt]
\noindent\textbf{Per Pettersson} is a senior researcher at the Norwegian Research Centre at the Energy Department. He is interested in hyperbolic problems, uncertainty quantification, numerical methods for subsurface CO2 storage, boundary conditions, and time-stability. Pettersson holds a double PhD degree in scientific computing from the Uppsala University and in computational and mathematical engineering from Stanford University.\\[8pt]
\noindent\textbf{Iuliu Sorin Pop} is a professor at University of Hasselt. He is interested in flow and reactive transport in porous media, biofilm growth, and geothermal energy. Pop holds a PhD degree in mathematics from the Babes-Bolyai University.\\[8pt]
\noindent\textbf{Kundan Kumar} is a senior lecturer of mathematics at Karlstad University, Sweden and holds an associate professor II position at the University of Bergen, Norway. He is interested in upscaled models and computational tools for coupled multiphysics processes, domain decomposition techniques, and iterative and multi-rate algorithms for coupled flow and geomechanics. Kundan holds a PhD degree (cum laude) in applied mathematics from the Eindhoven University of Technology, The Netherlands. \\[8pt]
\noindent\textbf{Florin Adrian Radu} is a professor of applied mathematics at University of Bergen. He is interested in mathematical modeling, analysis, and numerical simulation of multiphase flow and multicomponent reactive transport in porous media. Radu holds a PhD degree in mathematics and a habilitation degree from the University of Erlangen-Nuremberg, Germany. 
\end{document}